\documentclass[sigplan,screen]{acmart}
\usepackage{subfig}
\usepackage{graphicx} 
\usepackage{lipsum}
\usepackage{enumitem}

\copyrightyear{2024}
\acmYear{2024}
\setcopyright{none}
\acmConference[ASPLOS '24]{29th ACM International Conference on Architectural Support for Programming Languages and Operating Systems, Volume 1}{April 27-May 1, 2024}{La Jolla, CA, USA}

\usepackage[]{hyperref}

\author{Narges Alavisamani$^*$}
\affiliation{%
  \institution{Georgia Tech}
  \country{Atlanta, Georgia, USA}
}

\author{Suhas Vittal}
\affiliation{%
  \institution{Georgia Tech}
  \country{Atlanta, Georgia, USA}
}

\author{Ramin Ayanzadeh}
\affiliation{%
  \institution{Georgia Tech}
  \country{Atlanta, Georgia, USA}
}

\author{Poulami Das}
\affiliation{%
  \institution{University of Texas at Austin}
  \country{Austin, Texas, USA}
}

\author{Moinuddin Qureshi}
\affiliation{%
  \institution{Georgia Tech}
  \country{Atlanta, Georgia, USA}
}

\date{March 2024}

\renewcommand{\shortauthors}{\empty}

\def \methodName{Promatch}

\long\def\ignore#1{}

\usepackage[normalem]{ulem}

\usepackage{adjustbox}
\usepackage{stfloats}
\usepackage{amsmath,amsfonts}
\usepackage{xcolor}
\usepackage{tcolorbox}
\usepackage{algpseudocode}
\usepackage{algorithm}
\usepackage[flushleft]{threeparttable}
\begin{abstract}

    Fault-tolerant quantum computing relies on {\em Quantum Error Correction (QEC)}, which encodes logical qubits into data and parity qubits. Error decoding is the process of translating the measured parity bits into types and locations of errors. 
    To prevent a backlog of errors,  error decoding must be performed in {\em real-time} (i.e., within $1\mu s$ on superconducting machines).  {\em Minimum Weight Perfect Matching (MWPM)} is an accurate decoding algorithm for surface code, and recent research has demonstrated real-time implementations of MWPM (RT-MWPM) for a distance of up to 9.  Unfortunately, beyond d=9, the number of flipped parity bits in the syndrome, referred to as the Hamming weight of the syndrome, exceeds the capabilities of existing RT-MWPM decoders. In this work, our goal is to enable larger distance RT-MWPM decoders by using adaptive predecoding that converts high Hamming weight syndromes into low Hamming weight syndromes, which are accurately decoded by the RT-MWPM decoder.
        
    An effective predecoder must balance both accuracy (as any erroneous decoding by the predecoder contributes to the overall Logical Error Rate, termed as LER) and coverage (as the predecoder must ensure that the hamming weight of the syndrome is within the capability of the final decoder).  In this paper, we propose \methodName{}, a real-time adaptive predecoder that predecodes both simple and complex patterns using a locality-aware, greedy approach. Our approach ensures two crucial factors: 1) high accuracy in prematching flipped bits, ensuring that the decoding accuracy is not hampered by the predecoder, and 2)~ enough coverage adjusted based on the main decoder's capability given the time constraints. \methodName{} represents the first real-time decoding framework capable of decoding surface codes of distances 11 and 13, achieving an LER of $2.6\times 10^{-14}$ for distance 13. Moreover, we demonstrate that running \methodName{} concurrently with the recently proposed Astrea-G achieves LER equivalent to MWPM LER, $3.4\times10^{-15}$, for distance~13, representing the first real-time accurate decoder for up-to a distance of 13.

    \end{abstract}
\begin{document}

\title[\empty]{
    \methodName{}: 
    Extending the Reach of Real-Time Quantum Error Correction with Adaptive Predecoding
    }

\date{}
\maketitle
\pagestyle{plain}

\thispagestyle{empty}

\section{Introduction}

The inherent noisiness of quantum computers limits the execution of many promising applications in quantum chemistry, cryptanalysis, and machine learning~\cite{ childs2018firstqsim, gidney2021rsafactorization, kivlichan2020qsimelectrons, lee2021qchemtensorcontract, peruzzo2014variational, reiher2017nitrogenfixation, shor1999polynomial,huang2021information, huang2022quantum, wang2022quantumnas, anagolum2024eliv}. Furthermore, many of these applications require extremely low error rates ($<10^{-12}$) unlikely to be achieved on physical devices. \textit{Quantum Error Correction (QEC)} can enable these applications by forming fault-tolerant \textit{logical qubits} from multiple physical qubits~\cite{calderbank1996good, dennis2001surfacecodes, fowler2012surface, kitaev1997toriccodes, landahl2011colorcodes, shor1995scheme, steane1996multiple}. These logical qubits have lower error rates than their constituent physical qubits, and by increasing the level of redundancy, or \textit{code distance} ($d$), the logical error rate (LER) can be reduced.  Logical qubits are encoded using a combination of \textit{data} qubits, which encode a quantum state, and \textit{parity} qubits, which detect errors on the data qubits~\cite{calderbank1996good, dennis2001surfacecodes, fowler2012surface, kitaev1997toriccodes, landahl2011colorcodes, shor1995scheme, steane1996multiple}. To identify any errors on the data qubits, \textit{Fault-Tolerant Quantum Computers (FTQCs)} that use Quantum Error Correction (QEC) periodically execute syndrome extraction. This process involves executing a quantum circuit that measures the parity qubits. A measurement result of `$1$' indicates a parity-check failure. The parity qubit measurements are then compiled into a bitstring known as a \textit{syndrome}, which is subsequently sent to a classical \textit{decoder}. The decoder analyzes the syndrome to identify where errors have occurred on the logical qubit and computes a correction, which is sent to the control processor to update future operations. However, as the syndrome extraction operations are themselves faulty, decoders must analyze syndromes across $d$ rounds to accurately determine the location of errors.

\begin{figure*}[thb]
	\centering
	\includegraphics[width=0.98\textwidth]{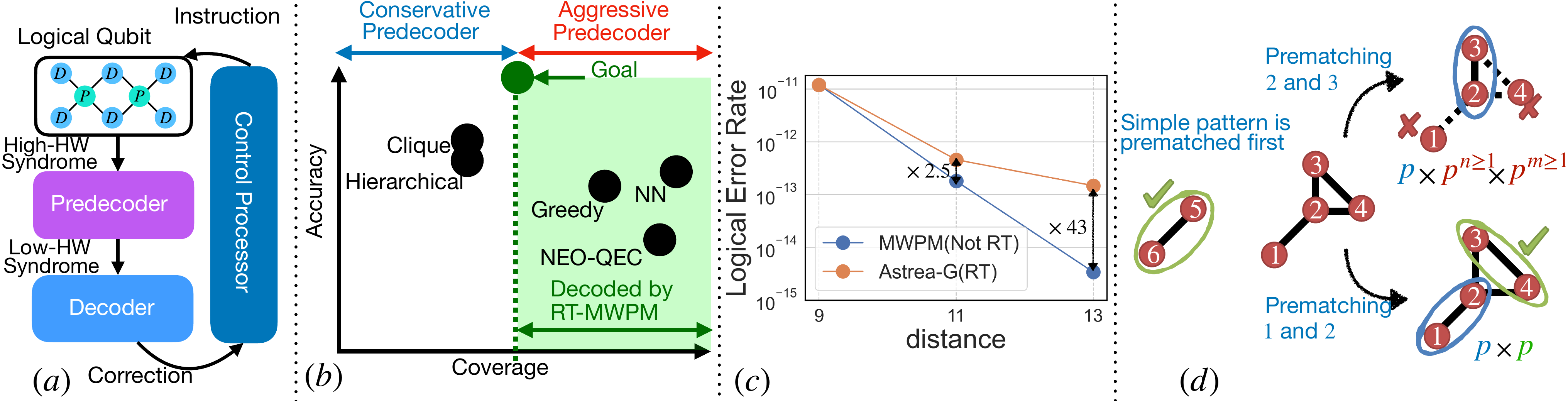}
    \caption{
		(a) The overview of QEC process in the presence of a predecoder. (b) The tradeoff between accuracy and coverage for Hierarchical predecoer~\cite{delfosse2020hierarchical}, Clique~\cite{ravi2023better}, and the greedy predecoder~\cite{smith2023local}. Existing predecoders either have low accuracy or low coverage.  (c) The gap between current RT-MWPM and Non-RT MWPM decoders indicating a 43x higher LER (d) An overview of insights using in \methodName{} for the locality-aware greedy algorithm. Promatch checks the neighborhood of flipped parity bits and finds the correct matching (nodes 1 and 2), which enables additional correct prematching (nodes 3 and 4). These two prematchings result in a lower weight, having a higher probability ($P^2$ compared to at most $P^3$).
	}
	\label{fig:intro_fig}
\end{figure*}

In recent years, several software-based implementations of MWPM have been proposed based on Blossom algorithm~\cite{higgott2023sparse, wu2023fusion, higgott2022pymatching, kolmogorov2009blossom} aiming to accelerate decoding and reduce average decoding time. However, they remain impractical due to the complexity of Blossom algorithm, which fails to guarantee $1 \mu s$ window. To achieve the required decoding speed, the decoding algorithms need to be implementable on hardware, such as FPGA, which serves as the primary focus of this work. Astrea~\cite{vittal2023astrea}, the state-of-the-art real-time MWPM (RT-MWPM) decoder, proposes using a brute-force method for decoding syndromes that has  up to 10 flipped parity bits. Number of flipped parity bits in the syndrome is referred to as \emph{Hamming weight (HW)} of the syndrome. Astrea is limitted to HW$\leq 10$ due to the exponential growth of number of matchings when HW increases. This fast growth makes applying brute-force method impossible for $d>7$. To address this challenge, a \emph{predecoding} method can be applied to predecode syndromes, accordingly reduce the HW of syndromes, and then send them to the main decoder, as shown in Figure~\ref{fig:intro_fig}(a). To have an effective predecoding process, the predecoder needs to 1) be accurate, not limiting the accuracy of the decoder 2) sufficiently cover enough number of flipped bits, making sure that the rest is manageable by the decoder. The goal of this paper is to develop a larger distance RT-MWPM using an adaptive predecoder that is accurate and has enough coverage.

In the context of predecoding methods, accuracy refers to the correctness of the predecoding process in matching flipped bits before they are passed to the main decoder. If predecoding is inaccurate, it results in erroneous decoding even if the main decoder is capable of accurately handling the remaining bits. Consequently, achieving high accuracy in predecoding is essential to ensure the reliability of the entire error correction process. In addition to accuracy, coverage is another important factor for predecoders, which refers to the number of pairs of flipped bits that are  predecoded compared to the total number of pairs required to be decoded. Insufficient coverage means that too many error pairs are left for the main decoder which cannot be handled within the time constraints. On the other hand, excessively high coverage underutilizes the main decoder's capability and may potentially compromise accuracy.

Prior predecoders tend to prioritize either coverage or accuracy~\cite{ravi2023better,delfosse2020hierarchical,smith2023local}, as shown in Figure~\ref{fig:intro_fig}(b). For example, Clique~\cite{ravi2023better} and Hierarchical decoder~\cite{delfosse2020hierarchical} which uses a simple predecoding structures for simple patterns and using MWPM for complex patterns. Therefore, these predecoders do not reduce the complexity requirement of the main decoder, and they are still reliant on having a main decoder that can decode high Hamming weight syndromes, which is impractical to perform in real-time. Furthermore, any inaccuracy of the predecoder still contributes to the overall logical error-rate.  On the other hand, Smith et al.'s work~\cite{smith2023local} proposes a Greedy predecoder that has high coverage but low accuracy.  Therefore, the overall logical error rate of the combined decoder is more than two orders of magnitude higher than the MWPM decoder alone. Another work Astrea-G~\cite{vittal2023astrea} adopts a greedy method that matches bit flips until the main decoder can handle the remaining decoding, guaranteeing sufficient coverage, which leads to similar LER as MWPM LER for distance 9 but reduced accuracy for higher distances ($43\times$ higher LER at $d=13$), as shown in Figure~\ref{fig:intro_fig}(c). The tradeoff between accuracy and coverage makes it challenging to achieve an ideal solution. Low coverage may lead to the main decoder being unable to accurately decode the remaining bits. Conversely, high coverage might utilize the main decoder's capacity inefficiently, but could come at the cost of reduced overall accuracy due to incorrect matches. \methodName{} strikes the right balance between accuracy and coverage ensuring precise decoding and efficient utilization of the main decoder's capabilities.

\methodName{} utilizes a locality-aware greedy algorithm, carefully considering the consequences of each decision made during the predecoding process. It matches and removes the flipped bits from the syndrome until the main decoder can find the exact MWPM solution for the remaining flipped bits before reaching $1 \mu s$.  In some cases, a group of flipped bits may form a complex pattern that is not immediately apparent how to match efficiently. However, \methodName{} utilizes the insight that by making \emph{the right initial matching} within a complex pattern, it can be broken down into simpler, more manageable patterns, as shown in Figure~\ref{fig:intro_fig}(d).

\methodName{} can be added to any decoder, however for our evaluations and getting low enough LER, we use Astrea~\cite{vittal2023astrea}. Our design incorporates Astrea's exact MWPM solution for low-HW syndromes and employs an adaptive matching approach to predecode flipped bits of the syndrome till reduce the HW to a value that allows us to apply Astrea within the remaining time before reaching $1 \mu s$. \methodName{} decodes surface codes of $d = 11$ and $d= 13$ in real time with LER of $4.5\times 10^{-13}$, and $2.6\times 10^{-14}$, respectively. The LER gained by \methodName{}, is the lowest LER in the literature that a real-time decoding process has achieved for distance 13. Further, we observed the gap between these LER values and MWPM LER can be closed by running Astrea-G~\cite{vittal2023astrea} in parallel with \methodName{}. Thus, our combined proposal represents the first decoder proposal that is both accurate (similar to MWPM) and real-time for up-to distance of 13. 

\vspace{0.05 in}

Overall, this paper makes the following contribution:
\vspace{0.05 in}

\begin{itemize}
    \item \emph{Accurate Predecoder}: Promatch introduces a locality-aware greedy method that utilizes information from each flipped bit's neighborhood to provide an adequate number of accurate matching while avoiding inaccurate patterns.

    \item \emph{Adaptive Predecoder:} Promatch is an adaptive predecoder which increases the complexity of patterns that are pre-matched based on whether the resulting hamming weight is within the capability of the main decoder. 

    \item \emph{Real-time Decoding for Large Distances:} \methodName{} is the first real-time decoder capable of decoding surface codes of distance 11 and 13 with the highest accuracy in the literature.

    \item \emph{Achieving MWPM LER for Large Distances:} When run concurrently with Astrea-G, \methodName{} makes it possible to achieve MWPM LER for up to $d=13$ for the first time.
\end{itemize}

\begin{figure*}[t]
	\centering
	\includegraphics[width=\textwidth]{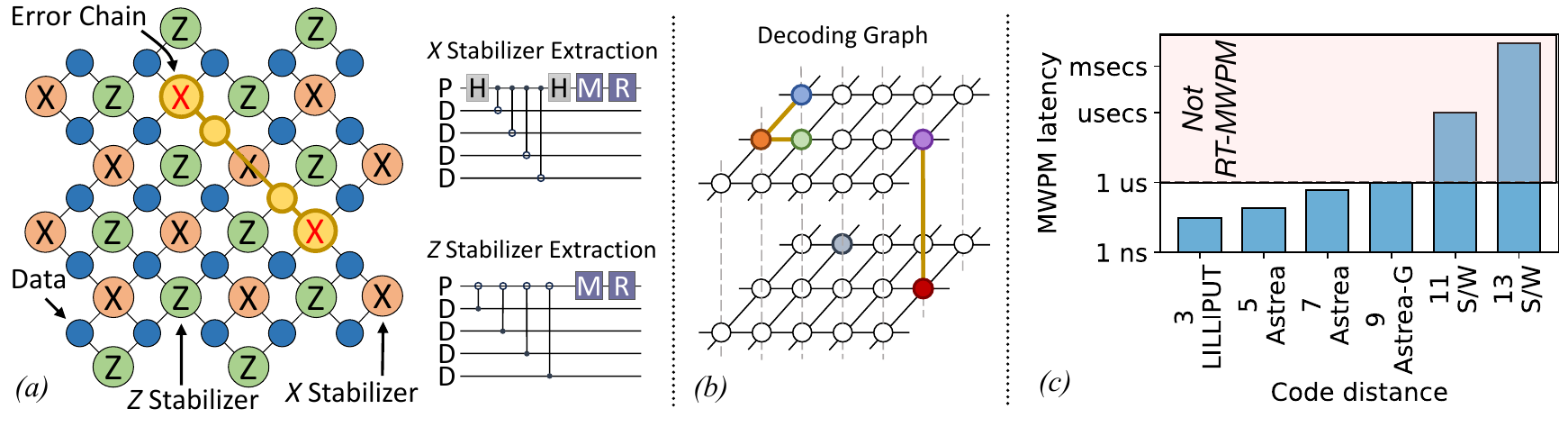}
    \caption{				
(a) Illustration of a $d=5$ surface code logical qubit lattice, alongside the associated $Z$ and $X$ stabilizer extraction circuits.
(b) Two-round decoding graph example for the $X$ stabilizer. 
(c) Real-time decoders, under $1\mu s$, exist for up to distance 9 while, for higher distances, 11 and 13, we need to rely on software-based decoders which have high latency.
	}
	\label{fig:background}
\end{figure*}

\section{Background and Motivation}

\subsection{Quantum Error Correction Using Surface Codes}

\emph{Surface codes} are regarded as the most promising QEC code due to their high threshold (about 1\%) and grid structure. Surface codes encode a logical qubit of distance $d$ onto a lattice with $d^2$ data qubits and $d^2-1$ parity qubits~\cite{kitaev1997toriccodes, dennis2001surfacecodes, fowler2012surface, yu2014lookup}, as shown in Figure~\ref{fig:background}(a). Errors on data qubits are projected into Pauli errors on their adjacent parity qubits through the periodic measurement of these parity qubits. The output of this parity qubit measurement is called a syndrome, and the measurement process is known as syndrome extraction.
During this process, each parity qubit measures a four-qubit operator, called a stabilizer, involving its four neighboring data qubits and extracts information about errors on them. 
Surface codes use two types of stabilizers ($Z$ and $X$) to detect bit-flip ($X$) and phase-flip ($Z$) errors, respectively. Errors lead to failed parity checks producing \textit{non-zero} syndromes. The code distance is the measure of the redundancy of the code as well as its error-correcting capability. A distance $d$ code can correct all error chains of at most length $\left\lfloor \frac{d-1}{2} \right\rfloor$. 
\subsection{Error Decoding}
QEC uses decoders that analyze the syndromes to identify the location and type of errors by \textit{matching} or \textit{pairing} the non-zero syndrome bits or failed parity checks. The problem can be reduced to a matching problem on a two-dimensional graph, known as the \textit{decoding graph}, where each node denotes a parity qubit and each edge denotes a data qubit. The pairing step matches the non-zero nodes and assigns errors onto the data qubits corresponding to the edges connecting them, as illustrated in Figure~\ref{fig:background}(b). However, in reality, syndromes are imperfect due to gate and measurement errors that occur during syndrome extraction. These errors result in failed parity checks across consecutive syndrome extraction rounds. To tolerate these errors, the decoder analyzes $d$ consecutive syndromes, resulting in a matching problem on a three-dimensional graph, as shown in Figure~\ref{fig:background}(b). Decoders must accurately identify errors in real-time to prevent the accumulation of errors. The typical latency that must be met is about $1 \mu s$ on superconducting systems (which corresponds to the time it takes to extract a syndrome). 
\subsection{Minimum Weight Perfect Matching (MWPM)}
The MWPM decoder is widely regarded as the gold standard for decoding surface codes. Each edge on the decoding graph is associated with a \textit{weight} that denotes the \textit{probability of error} causing the two nodes connecting the edge to flip. Therefore, constructing a fully-connected weighted graph comprising of the non-zero syndrome bits and \textit{perfectly matching} the nodes such that the \textit{total weight is minimized} enables us to determine the highest probability error event. 
Although the accuracy of MWPM is desirable, implementing it in real-time is challenging owing to the complexities of the inherent Blossom algorithm used to compute the MWPM.

Recent works solve real-time MWPM (RT-MWPM) decoding using alternate approaches. As shown in Figure~\ref{fig:background}(c), LILLIPUT achieves RT-MWPM in 29 ns and 42 ns for $d$=3 and $d$=5 (for only two syndrome rounds) respectively using lookup tables~\cite{yu2014lookup, das2022lilliput}. However, the size of the tables grows exponentially with the distance, limiting its scalability.  Astrea achieves RT-MWPM up to $d$=7 within 456ns~\cite{vittal2023astrea}. As each error chain (irrespective of its length) only flips up to two syndrome bits, the \textit{Hamming weight} of the syndromes corresponding to correctable errors remain within $10$ for $d$=7. The number of possible matchings for syndromes of Hamming weight 10 is 945. Astrea searches through them using an accelerated brute-force search in hardware. However, brute-force search is not scalable for larger code distances as the Hamming weight increases due to the increased redundancies. Astrea-G extends Astrea by searching greedily and prioritizing certain matchings~\cite{vittal2023astrea}. It achieves RT-MWPM in $1 \mu s$ for up to $d$=9. However, this greediness causes inaccuracies beyond $d$=9 and the logical error rate of Astrea-G is higher than the idealized MWPM, for example 43$\times$ for 13. Beyond $d$=9, we must rely on software implementations~\cite{kolmogorov2009blossomv, higgott2021pymatching, higgott2023pymatching2} to achieve MWPM accuracy or approximate solutions (such as AFS decoder)~\cite{holmes2020nisqplus, huang2020weighteduf, ueno2021qecool, ueno2022qulatis, das2022afs, scruby2022numericaljit} that trade-off accuracy for real-time decoding. Although a recent variant of the \textit{Blossom} algorithm has significantly lower complexity compared to the original implementation, the worst-case latencies are still a few hundred microseconds to milliseconds. \textit{Ideally, we want to expand the reach of RT-MWPM beyond d=9. }

\subsection{Prior Works on Predecoding}

\begin{figure}[t]	
	\centering
	\includegraphics[width=0.9\columnwidth]{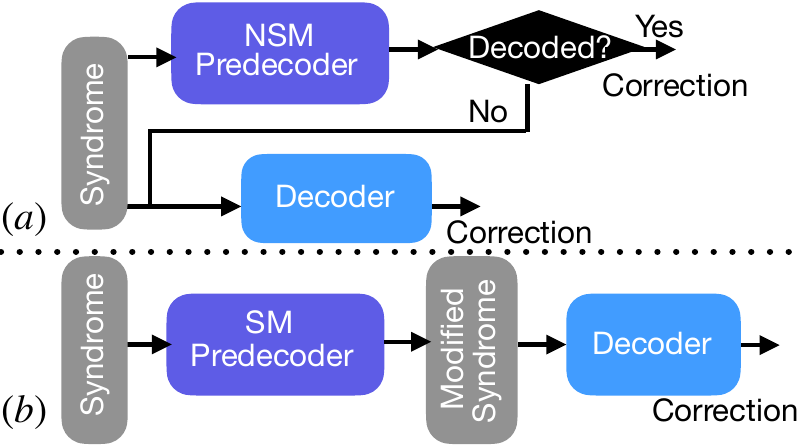}
    \caption{(a)~NSM Predecoders attempt to decode the entire syndrome. If they fail, then the entire syndrome is sent to the main decoder. (b)~SM Predecoders decode a subset of the syndrome and send the remainder to the main decoder.}	
    \vspace{-0.1in}
	\label{fig:predecdoer_types}
\end{figure}

The decoding complexity grows with the code distance due to the increased number of syndrome bit flips in the larger decoding graph. High Hamming weight syndromes are more complex and take longer to decode due to the increase in the number of possible pairings. The complexity of decoding can be reduced by \textit{prematching} or \textit{predecoding} a subset of the non-zero syndrome bits. The first proposals for predecoding focused on reducing syndrome transmission bandwidth between the quantum substrate and the decoders~\cite{delfosse2020hierarchical, ravi2023better}. These implementations attempt to quickly decode the entire syndrome by matching non-zero syndrome bits to their geometrically local neighbors. If the predecoder is able to do so, no information is sent to the main decoder (often an MWPM or Union-Find decoder). We refer to these predecoders as \textit{Non-Syndrome-Modified (NSM)} predecoders because they do not modify the syndrome before sending it to the main decoder, as shown in Figure~\ref{fig:predecdoer_types}(a).

More recent predecoders use an orthogonal approach that focuses on minimizing the complexity of the decoding task handled by the main decoder~\cite{caune2023belief, chamberland2022hierarchical, smith2023local, ueno2022neoqec}.  These \textit{Syndrome-Modified (SM)} predecoders reduce the Hamming weight of the syndromes by matching a subset of the syndrome bits and sending the remaining unmatched syndrome with a lower Hamming weight to the main decoder, as illustrated in Figure~\ref{fig:predecdoer_types}(b).


\subsection{Limitations of Prior Works on Predecoding}

The limitation of NSM predecoders is that these predecoders do not reduce the decoding complexity of the main decoder and therefore, the overall decoding performance is still constrained by the main decoder. If the main decoder is a software MWPM decoder, real-time decoding is not feasible~\cite{ravi2023better}, whereas the accuracy is reduced if the main decoder is a Union-Find decoder~\cite{delfosse2020hierarchical}. Concurrently, SM predecoders are limited by their accuracy: such predecoders may err when matching, and this error will cause a logical error\footnote{We note that the accuracy of any predecoder, both NSM and SM, is crucial as any inaccuracies will cause a logical error.}.

Thus, we observe a tradeoff between predecoder \textit{accuracy} and \textit{coverage}. NSM predecoders avoid predecoding syndromes with potential non-local matchings, resulting in subpar coverage. In contrast, Syndrome-Modification predecoders achieve good coverage by reducing the Hamming weight of the syndrome, but may incur inaccuracy while doing so. We further note that coverage is strongly correlated to \textit{latency}, and there is an inherent trade-off between accuracy and decoding latency. A predecoder that optimizes for accuracy will avoid reducing the Hamming weight too significantly so the main decoder decodes the majority of the syndrome. However, this increases overall decoding latency. 
For example, the Clique predecoder is limited by the latency of software MWPM~\cite{ravi2023better}. 
On the other hand, a predecoder that optimizes for coverage may predecode too much of the syndrome to relieve the burden of the main decoder. This approach reduces decoding latency but also reduces accuracy. 
This phenomenon is observed in the predecoders proposed by Chamberland et al. and Smith et al.~\cite{chamberland2022hierarchical, smith2023local}. 
Unfortunately, prior predecoders optimize for either higher accuracy or coverage. Note that it is possible to have very accurate and high coverage predecoding using complex algorithms such as belief propagation but incur long latencies that may not converge in real-time~\cite{caune2023belief}.

\subsection{Goal: Enabling Higher Distance RT-MWPM Decoders by Using Adaptive SM Predecoding}
\begin{figure}[h]	
	\centering
	\includegraphics[width=\columnwidth]{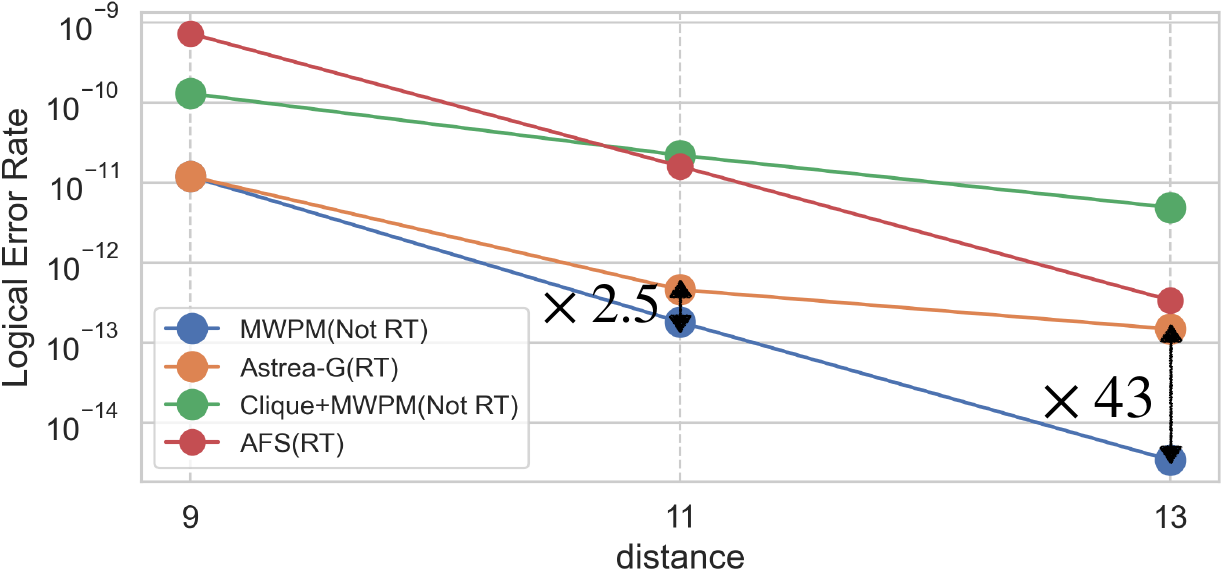}
    \caption{
		Logical error rate trends for MWPM, Astrea-G, Clique+MWPM, and AFS as code distance $d$ increases, considering a physical error rate of $10^{-4}$.
	}
	\label{fig:LER_trend}
\end{figure}

Currently, there is a gap between RT decoders and Non-RT MWPM for $d > 9$, $2.5$ times and $43$ times higher than logical error rate of Non-RT MWPM, as illustrated in Figure~\ref{fig:LER_trend}. Notably, a contemporary RT MWPM decoder, named Astrea, can decode syndromes that have low Hamming weight. Motivated by this capability, our objective is to construct a SM predecoder that can precisely predecode and reduce high Hamming weight syndromes. This reduction enables the RT MWPM decoder to process the modified syndromes effectively. Our envisioned SM predecoder operates adaptively, predecoding syndromes to achieve a level of \emph{sufficient} coverage based on the present abilities of the RT MWPM decoder. With this approach, we aim to design an accurate RT predecoder that, when combined with Astrea, can bridge the existing gap between RT and Non-RT MWPM decoders.

\section{\methodName{}: Key Insights}

\methodName{} focuses on decoding syndromes that have more than 10 flipped bits (HW $> 10$), which we refer to as high Hamming weight syndromes as Astrea can accurately decode all syndromes with HW $\leq 10$ in real-time. Like many other predecoding approaches, the core idea of this work revolves around the observation that most of the flipped bits in the syndrome are matched to their neighbors in the decoding graph (indicating a chain of length one).

\begin{figure}[h]	
	\centering
 \vspace{0.1 in}
	\includegraphics[width=\columnwidth]{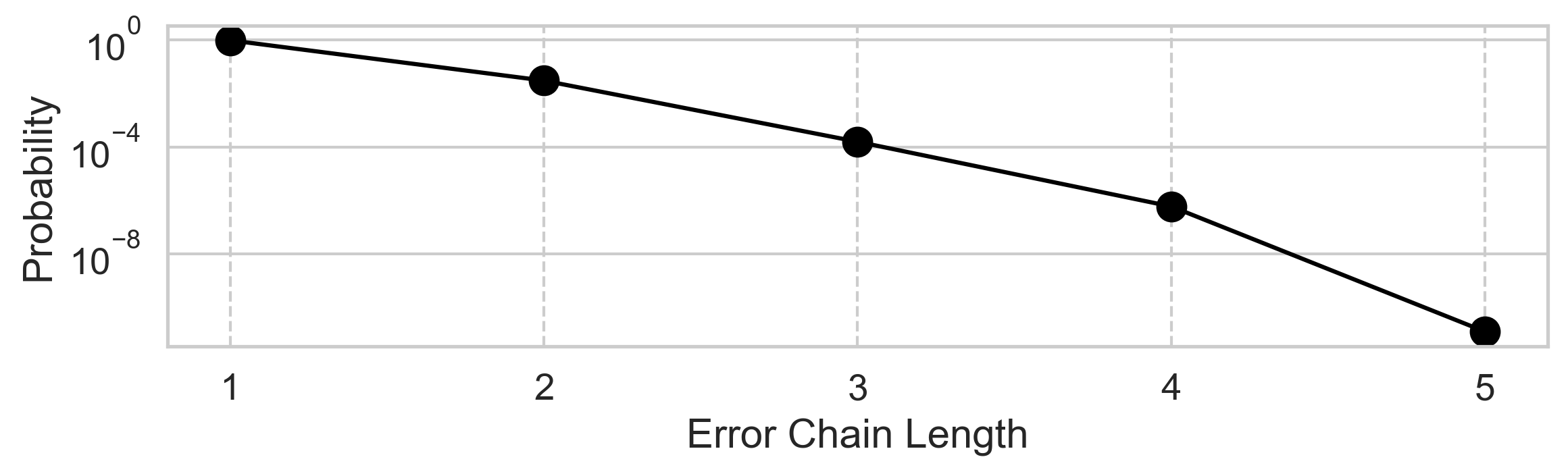}
    \caption{ More than $90\%$ of error chains, based on MWPM decoder, has length of 1. This means more than $90\%$ of flipped bits are matched to their neighbors. This plot is for distance 13 and physical error rate $10^{-4}$.
		 }	
	\label{fig:ec_distr}
\end{figure}

\begin{figure}[t]	
	\centering
	\includegraphics[width=\columnwidth]{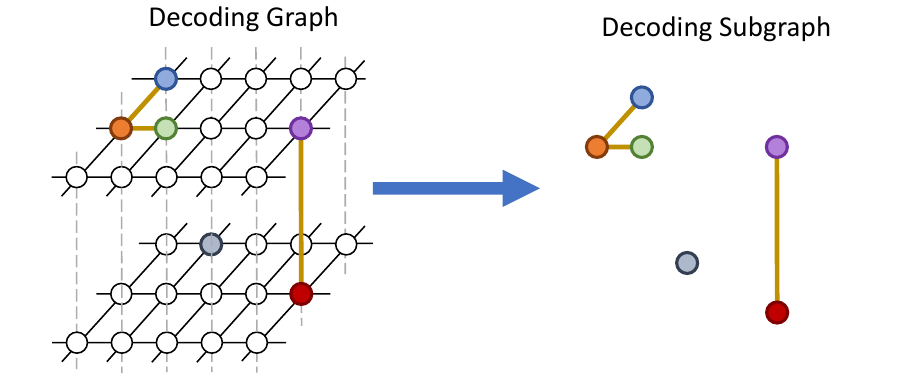}
    \caption{ Decoding subgraph created from flipped parity bits and edges of which both nodes (parity bits) are flipped.}	
	\label{fig:subgraph}
\end{figure}

Figure~\ref{fig:ec_distr} shows the frequency of different error chain lengths for the high Hamming weight syndromes. As error chains of length $1$ are extremely common, most predecoders attempt to remove such errors~\cite{delfosse2020hierarchical,  smith2023local}. However, only greedily predecoding errors via error chains of length 1 leads to a loss in accuracy. In this section, we present our insights regarding how incorrect decisions can be avoided during predecoding. In our insights, we leverage the \textit{decoding subgraph}, which is the subgraph of the larger decoding graph containing only the nonzero bits in a syndrome bit and any edges between them, as shown in Figure~\ref{fig:subgraph}.

\subsection{High Accuracy: Singletons contribute more weights to the MWPM solution}

During predecoding, every prematching removes two nonzero syndrome bits from the decoding subgraph, consequently removing any edges incident to these syndrome bits. As edges are removed from the graph, this can result in syndrome bits unconnected to the rest of the decoding subgraph. We call these unconnected bits \textit{singletons}. As these singletons are not connected to any other syndrome bit, the only way to match such bits to other syndrome bits is by correcting along an error chain with length $L \geq 2$. However, such error chains are unlikely, as their probability of occurrence is about $p^{-L}$. Thus, our first insight is that \textit{predecoding should minimize the number of singletons created during prematching}.

\begin{figure}[h]	
	\centering
	\includegraphics[width=0.68\columnwidth]{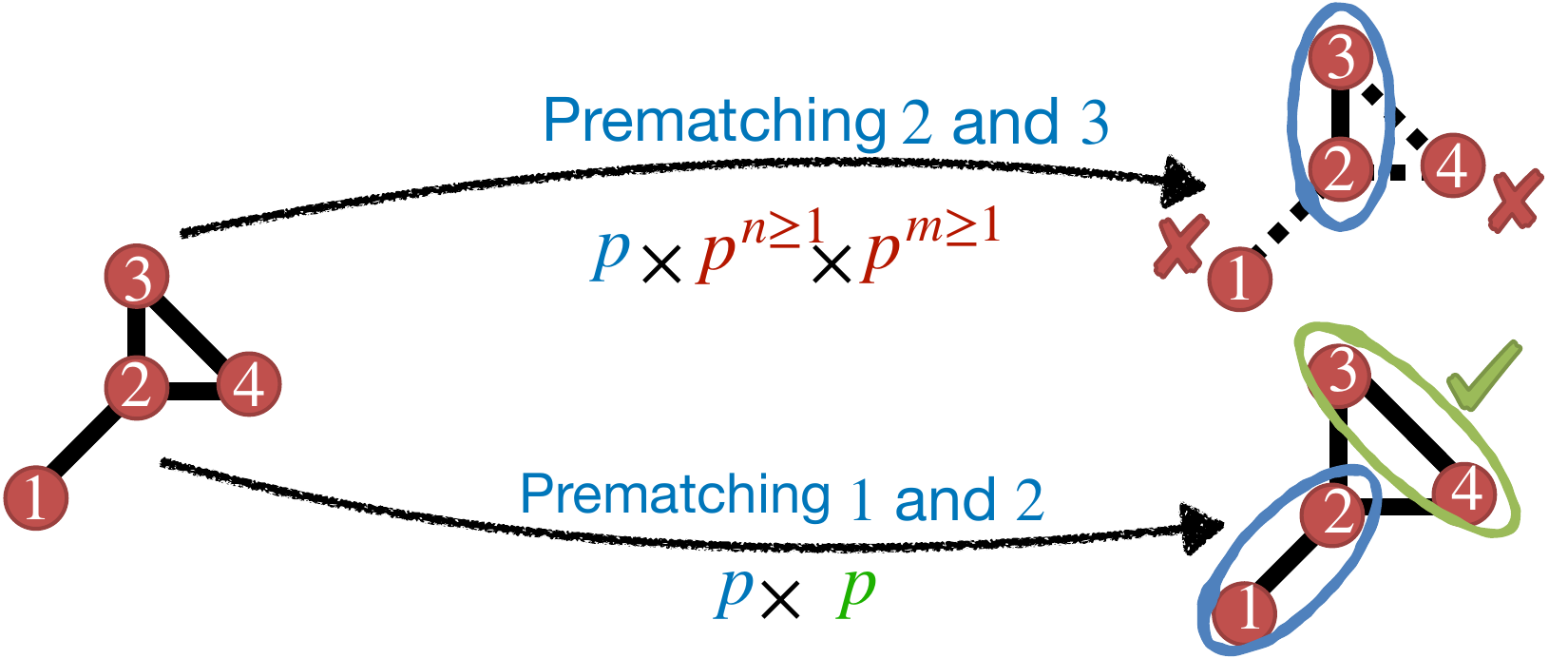}
    \caption{ Avoiding generation of singletons results in lower-cost matchings in future steps. Additionally, a correct matching (nodes 1 and 2) enables additional correct prematching (nodes 3 and 4).}	
     \vspace{-0.15in}
	\label{fig:insight}
\end{figure}

In Figure~\ref{fig:insight}, we illustrate the structure of four connected flipped bits observed in the simulations. In this scenario, we have six possible options for matching, and the correct choice is to match 1 with 2 and 3 with 4. However, if we make a mistake and match 2 with 3, in the next step, we will have two singletons of 1 and 4, and we will have no other option but to match them. This incorrect matching solution results in a total weight of approximately 12, whereas the correct match has a weight of around 8. It is important to note that the correct match is the only one that does not generate any new singletons, making it the optimal choice.

\begin{figure*}[t!]	
	\centering
	\includegraphics[width=\linewidth]{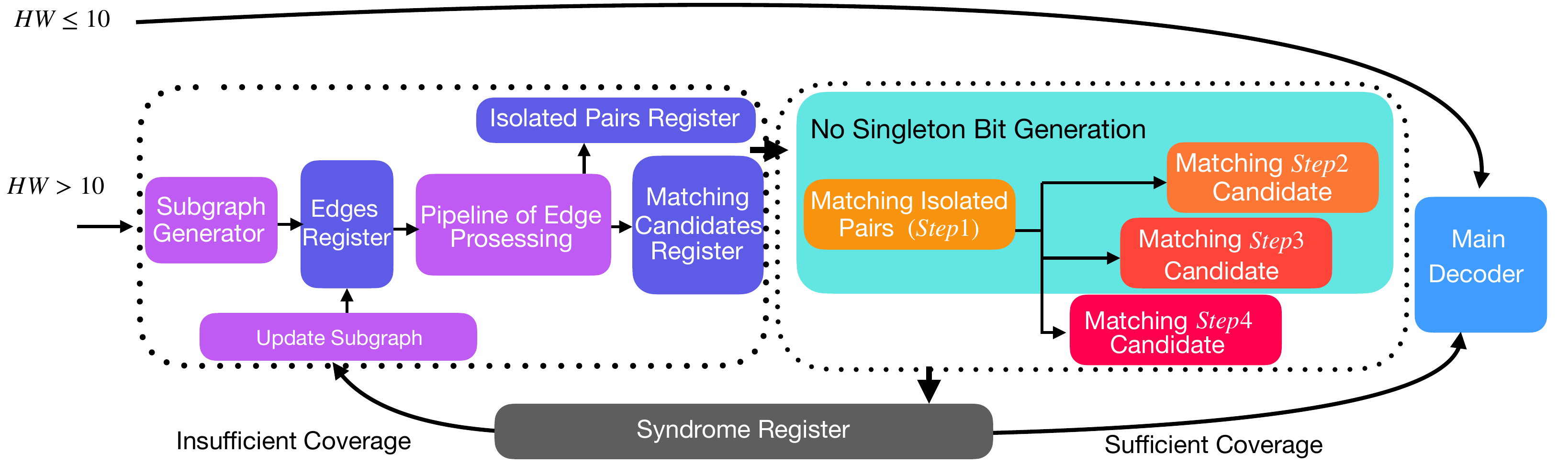}
    \caption{Overview of \methodName{}}
	\label{fig:overview}
\end{figure*}

\subsection{Sufficient Coverage: Use multiple simple steps}

Prior predecoders treat prematching as a monolithic process: the predecoder prematches once, and the remainder of the syndrome, regardless of the remaining Hamming weight, is sent to the main decoder. However, we observe that prematching decisions often reveal new possible prematchings by removing nodes and edges from the decoding subgraph and consequently reducing the complexity of the decoding subgraph. Thus, our second insight is that \textit{prematching decisions enable additional prematching decisions} \footnote{We assume independent errors (no spatial correlation). 
 If multiple errors happen far away from each other then such isolated errors can be matched easily, and decoding becomes an easy problem.  However, the harder problem for decoding is when multiple errors happen to occur near to each other and this occurs with non-negligible probability even with random errors. Handling such patterns is a requirement to achieve low logical error rate (for example, if such patterns happen with a 1 in a billion probability and the decoder is unable to  handle, then the LER will always be at-least $10^{-9}$, whereas we seek LER of as low as $10^{-15}$), so we need to handle such patterns. }

As shown in Figure~\ref{fig:insight}, the structure of four connected flipped bits may appear complex at first glance. However, by following the key insight of not generating new singletons, we can make the optimal matches. For instance, we match 1 with 2, leaving us with the simple match of 3 with 4 as a stand-alone pair. In this way, we can successfully match an error structure of four flipped bits in just two consecutive simple steps. This demonstrates the effectiveness of our approach in handling complex error patterns while minimizing the adverse impact of matching decisions.

\ignore{
\subsection{High Accuracy + Sufficient Coverage: A locality-aware greedy approach}
Not all options have equal consequences and matching weights. Some choices are less risky, having no destructive consequences and lower weights. This is particularly important because the predecoder does not need to match all the flipped bits. To achieve efficient error correction, prioritizing safe matching options is key. Using a locality-aware greedy method, we reduce the immediate matching weight and also factor in how current decisions might influence future ones, aiming to prevent creating singletons. Then, we move towards riskier options until achieving the required coverage, enabling effective utilization of predecoded syndromes by the main decoder. 
}
\ignore{Unlike many predecoders that ignore complex patterns, \methodName{} ensures sufficient matching to allow the remaining flipped bits to be decoded in the remaining time by the last-level decoder. This means that \methodName{} also decodes complex patterns, breaking them down into simpler patterns through proper matching. For instance, in Figure xx, a complex pattern initially seems challenging to match. However, by matching node 1 with node 2, nodes 3 and 4 can be trivially matched as well.
Additionally, not all syndromes require risky matching decisions, as many syndromes only consist of simple patterns. Building on these insights, we designed multiple stages for predecoding, where the first stages prioritize decoding simple, less risky matchings to achieve high accuracy. If necessary to achieve sufficient coverage, \methodName{} continues matching more complex and riskier patterns in later stages.

By decoding both simple and complex patterns strategically, \methodName{} effectively balances accuracy and coverage in the predecoding process. This approach ensures that the decoding is accurate and reliable for simpler patterns, while also handling more complex patterns to maintain sufficient coverage. Consequently, \methodName{}'s multi-stage predecoding method contributes to its ability to achieve both high accuracy and enough coverage, making it a robust solution for real-time surface code error decoding.

To ensure high accuracy, we devised a strategy that prioritizes matching two flipped bits in a way that avoids introducing new singleton bits and selects pairs with the highest probability of success (i.e., the lowest weight of the edge in the decoding subgraph) among those that don't generate singletons. By preventing the creation of new singleton bits, we effectively limit the additional weight added to the MWPM solution. As in our noise model, we use uniform physical error rates ($p = 10^{-4}$), making the weights of edges similar and the summation of the weights of any two edges (the multiplication of error rates associated with any two edges) is higher (lower) than the weight of any single edge (the error rate associated with that edge).

Based on this key insight, we designed the first three stages of our decoder. During these stages, \methodName{} exclusively matches flipped bits that are neighbors and have no other neighbors of degree $1$. This ensures that if a given node remains unmatched, its neighbors would become singleton bits as they have no other neighbor to match with.

With this approach, \methodName{} effectively prioritizes matching flipped bits in a manner that avoids creating singleton bits and maintains high accuracy throughout the initial stages of the decoding process. By selecting the most favorable pairs and considering the consequences of matching decisions, our method achieves both high accuracy and sufficient coverage, making it a powerful solution for real-time surface code error decoding.}

\section{\methodName{} Design}
 This section explains the algorithm and  hardware design of \methodName{}. Figure~\ref{fig:overview} shows an overview of \methodName{}. 
 \methodName{} is designed in such a way that it initiates the matching process with the least risky pairs. If necessary, it incrementally adjusts the risk level until a sufficient coverage is achieved.
It leverages the insight that initial matching of complex patterns into simpler ones is crucial. Specifically, during stages 2, 3, and 4, which tackle complex patterns, \methodName{} matches one pair at a time before reassessing. This approach can simplify patterns, allowing earlier stages to address them.
 For example, in Figure~\ref{fig:insight}, prematching bits 1 and 2 breaks the complex pattern which results in trivially matching 3 and 4.
After each matching that \methodName{} applies, it checks if the main decoder can decode the modified syndrome in the remaining time. If yes, it gives the syndrome to the main decoder. If not, \methodName{} predecodes more bits.

\subsection{\methodName{} Algorithm}\label{sec:algorithm_steps}
\begin{figure}[h]
	\centering	
	\includegraphics[width=\columnwidth]{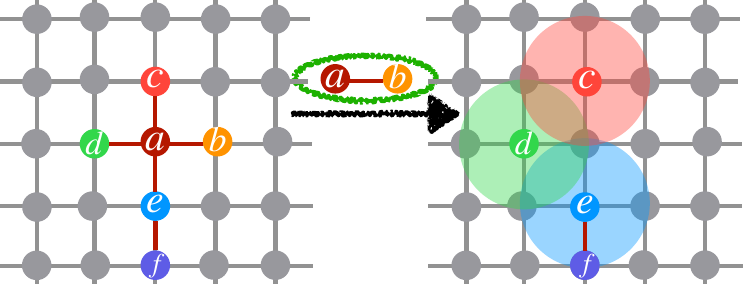}
    \caption{
        There are three nodes ($b$, $c$, and $d$) in the decoding subgraph, whose only neighbor is node $a$. If $a$ and $b$ get matched with each other, it creates singleton bits $c$ and $d$ (while $e$ pairs with $f$). 
	}    
	\label{fig:dependent_nodes}
\end{figure}

At each round of \methodName{}, we extract the data of the decoding subgraph. Each flipped bit that has not yet been matched in the syndrome acts as a node in the subgraph. We gather the following data for each node of the subgraph at each round of the algorithm: 1) The degree of the node. For node $i$, this is denoted by $deg_i$. 2)~For each node, the number of neighbors possessing degree 1.  For node $i$, this number is denoted by $\#dependent_i$. $\#dependent_i$ indicates the number of flipped bits in the syndrome that rely exclusively on node $i$ for a match with an error chain of length 1. If these neighboring bits do not get matched with node $i$, they become singletons. For instance, in Figure~\ref{fig:dependent_nodes}, flipped bit $a$ has four neighbors, $b$, $c$, $d$, and $e$. 
Three out of these four neighbors have a degree of 1, namely $b$, $c$, and $d$. This means that if these three nodes intend to form a match in an error chain of length 1, bit $a$ is their sole option.
For example, if $a$ and $b$ get matched and removed from the syndrome, the other \emph{degree-1} neighbors, $c$ and $d$, will have no neighbors in their neighborhood with only 1 edge and will become singleton bits. However, $e$ still has $f$ in its neighborhood, preventing it from becoming a singleton.

After updating all the $\#dependent_i$ and $deg_i$ values for all nodes, \methodName{} starts finding matchings with prioritizing pairs that: (a) do not generate singletons after matching, and (b) have the highest probability,  meaning they minimally increase the overall weight in the MWPM solution.
To ensure high accuracy, \methodName{} begins by matching {\em isolated pairs}—simple patterns of flipped bits with only one neighbor, which are more common and less risky. To achieve sufficient coverage, it also engages with complex patterns, involving flipped bits that have multiple neighbors and, consequently, multiple matching options. These patterns can also contain extant singletons, which are complicated by their lack of adjacent matching options. Despite these complex patterns being riskier and less frequent, \methodName{} dedicates part of its design to these patterns to ensure enough coverage. 
In the following parts, we describe the steps of \methodName{}, ordering them by their priorities. In addition, we explain the reasons behind their prioritization order. These steps are also elaborated in Algorithm~\ref{alg:cap}.
\begin{enumerate}
    \item \methodName{} begins by matching isolated pairs primarily for two key reasons: First, based on our insight about how singletons can increase the weight of the MWPM solution, \methodName{} prioritizes matching isolated pairs because matching them does not create singletons and accordingly ensures that the decoding process remains efficient by not introducing higher weights. Second, if any of the flipped bits among the two of bits in an isolated pairs gets matched to another node, the other node becomes a singleton bit. Therefore, to prevent either of flipped bits in an isolated pair from becoming a singleton bit, we require to match flipped bits of an isolated pairs with each other.
    
    \item In this step, \methodName{} matches two neighboring flipped parity bits, that creates a non-isolated pairs, only if the matching does not generate any singleton bits.
        \begin{enumerate}[label*=\arabic*.]
            \item First, \methodName{} prioritizes a  pairs of neighboring nodes with the lowest weight in the decoding graph (highest probability). In this step, it prioritizes the pair of which one of the nodes have degree~1. Similar to step~1, it is important to prioritizing matching such pair because for one of the bits, this matching is the only option that prevents it from becoming a singleton bit.
            \item Second, if there is no pair of which the degree of the nodes is~1, it chooses the pair that has the lowest weight in the decoding graph (highest probability).
        \end{enumerate}
        
        \item \methodName{} employs this particular step only when there are no viable matching candidates left for step~2. This situation arises when pairs of neighboring flipped bits cannot be matched without leading to the creation of a singleton bit. In such cases, \methodName{} opts to match an extant singleton bit with another flipped bit, choosing the one that forms the shortest path or, equivalently, the error chain with the highest probability. In this step, the condition of not creating singleton bit is still necessary. it is important to note that, \methodName{} requires to search among a fewer number of paths, compared
to Astrea’s brute-force method. This is because the number of singleton bits is low. Additionally, this step is utilized only after Step 1 and 2 have been applied and if Step 1 and Step 2 have not been enough to reach the sufficient coverage. Therefore, there are a few number of paths left, which makes this step fast enough.

        \item The algorithm uses this step only if no flipped bit can be matched in Step~1, Step~2, and Step~3. This step is similar to Step~2, having similar substeps~4.1 and 4.2, without the condition of not generating singleton bits, and is the only one that adds singleton bits to the decoding subgraph. Therefore, in this step, \methodName{} takes the riskiest decision to ensure that the main decoder can decode the modified syndrome within remaining time before reaching $1 \mu s$. Note that, similar to Step~2 and Step~3, \methodName{} may break complex patterns to simple patterns allowing  less risky decisions in future.  

\end{enumerate}

\begin{algorithm}
\caption{\methodName{} Algorithm}\label{alg:cap}
\begin{algorithmic}
\State {\textbf{Input: }Decoding Subgraph $(V,E)$}
\While{HW is \textbf{not} low enough}
    \State {\textbf{while} isolated pairs exist \textbf{and} HW is not low enoug \textbf{do}}
    \State{\quad\quad match isolated pairs}  
    \State {\textbf{if} HW is low enough \textbf{then} break}
    \State{\textbf{for} $e_{ij}$ in $E$}
        \State {\text{\quad\quad \textbf{if}} matching $i$ and $j$ does not create Singleton} 

            \State {\text{\quad\quad\quad \textbf{if}} $min(deg_i,deg_j) == 1$}
                    \State {\text{\quad\quad\quad\quad}\textbf{if} $w_{ij} <$ weight of current \emph{Step2.1} candidate}
                    \State{\text{\quad\quad\quad\quad\quad} \textbf{then} set $(i,j)$ as \emph{Step2.1} candidate}
            \State {\text{\quad\quad\quad \textbf{else if}} $w_{ij} <$ weight of current \emph{Step2.2} candidate}
            \State{\text{\quad\quad\quad\quad} \textbf{then} set $(i,j)$ as \emph{Step2.2} candidate}
        \State {\text{\quad\quad \textbf{else}}} \Comment{Risky step}
            \State {\text{\quad\quad\quad \textbf{if}} $min(deg_i,deg_j) == 1$}
                    \State {\text{\quad\quad\quad\quad}\textbf{if} $w_{ij} <$ weight of current \emph{Step4.1} candidate}
                    \State{\text{\quad\quad\quad\quad\quad} \textbf{then} set $(i,j)$ as \emph{Step4.2} candidate}
            \State {\text{\quad\quad\quad \textbf{else if}} $w_{ij} <$ weight of current \emph{Step4.2} candidate}
            \State{\text{\quad\quad\quad\quad} \textbf{then} set $(i,j)$ as \emph{Step4.2} candidate}
        \State {\text{\quad\quad \textbf{end if}}}
    \State{\textbf{end for} }
    \State {\text{\textbf{if}} \emph{Step2.1} and \emph{Step2.2} candidate is empty }
    \State {\text{\textbf{and}} $\exists$ Singleton $\in V$ }
    \State{\quad\textbf{for} every node ${i}$ and Singleton node $j$ in $V$}
        \State {\text{\quad\quad \textbf{if}} matching $i$ and $j$ does not create new Singleton} 
        \State {\text{\quad\quad\quad \textbf{and} path weight of $i$ and $j$ is less than path}}
        \State {\text{\quad\quad\quad\quad weight of \emph{Step3} candidate}}
        \State {\text{\quad\quad\quad\quad\quad \textbf{then} set $(i,j)$ as \emph{Step3} candidate}}
        \State {\text{\quad\quad \textbf{end if}}}
    \State{\quad\textbf{end for} }
      \State{\textbf{end if} }
      \State{Match only 1 pair among the candidates prioritizing}
      \State{\emph{Step2.1}, \emph{Step2.2}, \emph{Step3}, \emph{Step4.1}, and then \emph{Step4.2}.}

\EndWhile
\end{algorithmic}
\end{algorithm}

\subsection{Hardware Implementation of \methodName{}}
\methodName{} iterates over the edges of decoding subgraph. The weights of the edges of the decoding graph are stored in a Edge Table in an on-chip memory on the FPGA.  To avoid impacting memory latency on the decoding time, the data is loaded from the memory gradually while the syndrome is extracted and previous syndrome is being decoded. Note that \methodName{}'s locality-aware approach makes loading all the required data during the syndrome extraction possible, which is not the case for exhaustive search approaches. There are four steps in \methodName{} algorithm design, detailed in Section~\ref{sec:algorithm_steps}. Each of these steps has its own matching candidates. The candidates of Step~1, isolated pairs, are stored in a separate register from the rest of the candidates because this step can have multiple candidates. This setup allows all these candidates to be applied to the syndrome simultaneously, a notable difference from other steps that match and modify the syndrome based on a single pair at a time. The candidates of other steps are stored in Matching Candidate Register. 

\subsubsection{Implementation of the Decoding Subgraph} The decoding subgraph information is stored in this format: A vertex array is utilized to store the index of the flipped parity bits in the syndrome. A neighbor array is also utilized to specify the neighbors of each vertex. Each neighbor contains the weight of the edge connecting it to the vertex. Additionally, each vertex $i$ has two vertex property array, namely the degree array and the \emph{dependency} array. which contains $deg_i$ and $\#dependent_i$, respectively. 

\subsubsection{The Pipeline of Finding Matching Candidates For Each Step}
\begin{figure}[t]
	\centering	
	\includegraphics[width=\columnwidth]{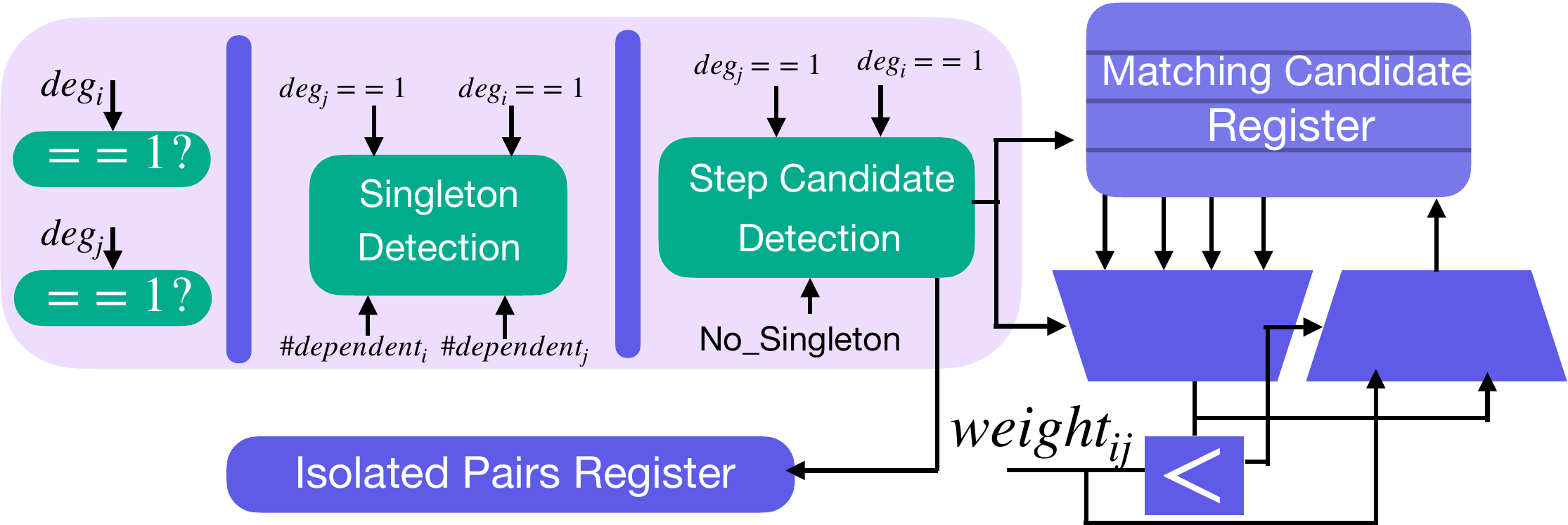}
    \caption{
	Pipeline of iterating over the edges of subgraph to find matching candidates for \methodName{}.}    
	\label{fig:edge_processing_pipeline}
\end{figure}

Figure~\ref{fig:edge_processing_pipeline} depicts the pipeline stages for identifying matching candidates within the decoding subgraph. The first stage examines each edge's node degrees, signaling which of the four steps in \methodName{} could consider the edge as a matching candidate. The subsequent stage evaluates if pairing through this edge would result in a singleton bit; if it does not, the edge is flagged as a potential candidate for Step~2 of \methodName{}. The third stage determines the appropriate candidate register for updates based on this edge. Lastly, it compares this edge's weights with the existing candidates, updating the candidate register if this edge offers a lower weight ( higher probability match).

\begin{figure}[htb]
	\centering	
	\includegraphics[width=1\columnwidth]{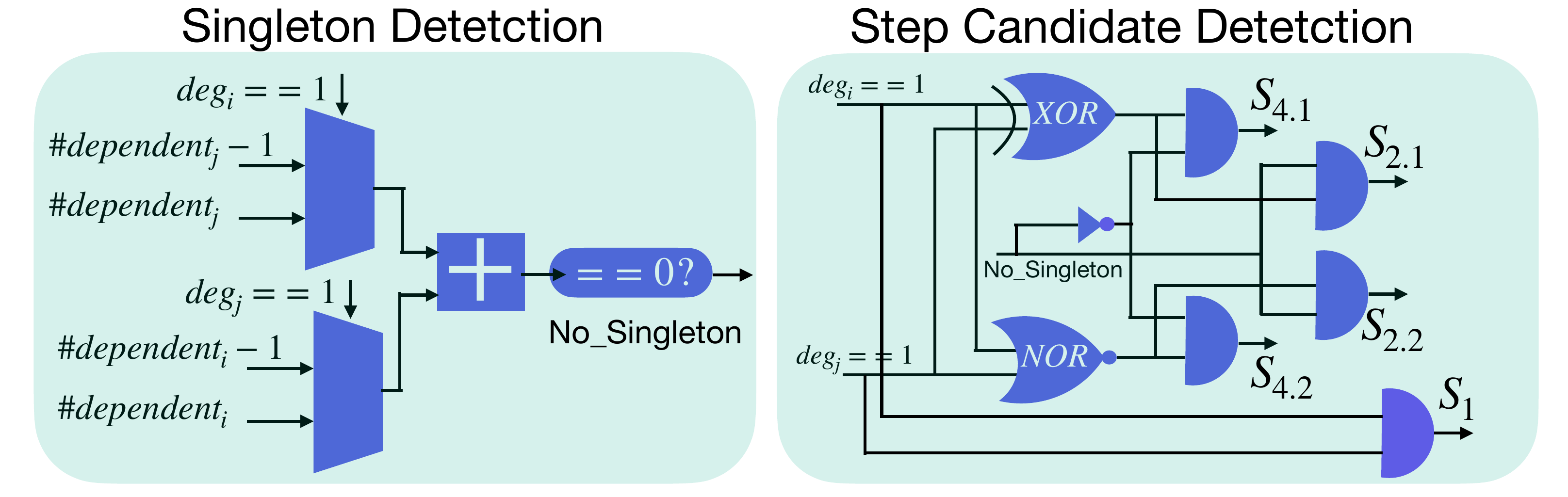}
    \caption{
	Logic for Singleton Selection and Step Selection of \methodName{}'s Algorithm.}    
	\label{fig:step_selection}
\end{figure}

Figure~\ref{fig:step_selection} shows the simple logics for singleton detection and step candidate detection, which makes \methodName{} fast enough compatible with the time limitation that the decoding process has.

In Step 3, \methodName{} calculates the lowest-weight paths between existing singleton bits and the remaining flipped bits. It uses an $ n \times n$ weight table, with each cell containing 8 bits, to store path weights among all 
$n$ syndrome bits. This table, kept in on-chip memory, is processed in parallel to and independently from the main pipeline of Figure~\ref{fig:edge_processing_pipeline}.

\subsubsection{\methodName{} || Astrea-G}

\methodName{} demonstrates a high proficiency in decoding a range of syndrome patterns, from simple to complex, prioritizing simpler matchings of isolated pairs or nodes with a single degree. This approach, however, faces challenges with more intricate error structures, especially those involving multiple distinct components located comparably close to each other, resulting in having multiple closely good matching options. AstreaG, on the other hand, use MWPM graph  which is a complete graph comprising all flipped parity bits. In this graph, the edges represent the shortest path between each pair of flipped bits. AstreaG filters out higher weight pairs by pruning edges of the MWPM graph with error chain probabilities below the LER, followed by employing a greedy-based near-exhaustive search method for decoding. When run in parallel, Astrea-G helps \methodName{} by providing an exhaustive search strategy. \methodName{} and Astrea-G each have their specific strengths in handling different types of error patterns which we explain in the following parts.




\textbf{Both Succeed in Sparse Error Patterns:} Both \methodName{} and Astrea-G are successful in decoding scenarios with sparse flipped parity bits. \methodName{}'s greedy approach works well in these cases, as it efficiently prioritizes isolated pairs or nodes with a single degree. Astrea-G also performs effectively here, as its searching method benefits from the sparsity, as it can prune more number of edges allowing for rapid convergence.

\textbf{\methodName{} Succeeds, Astrea-G Struggles:} In situations where decoding subgraph components are closely spaced but has just enough number of simple matchings, such as isolated pairs, \methodName{} outperforms Astrea-G. Astrea-G's exhaustive search method becomes less effective under real-time constraints in these dense environments. It struggles prune the MWPM graph. Therefore, it cannot effectively navigate through the tightly packed, yet non-interconnected components within the strict time limit.

\begin{figure}[h]
	\captionsetup[subfigure]{position=top} 
	\centering	
	\subfloat[]{
		\includegraphics[width=0.5\columnwidth]{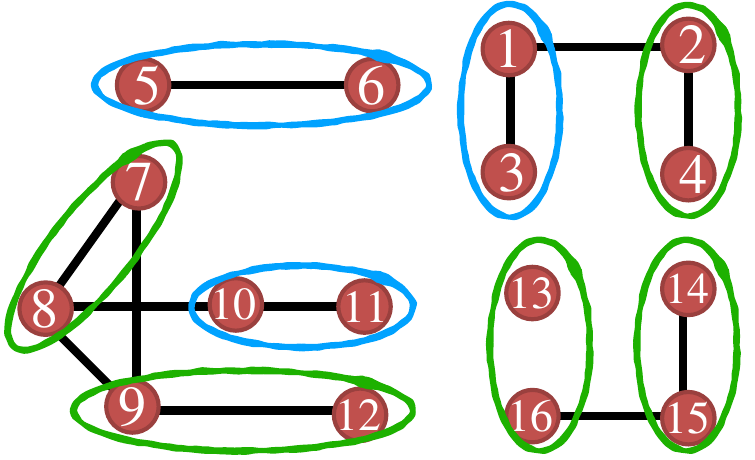}
	}\hspace*{-0.8em}		
	\subfloat[]{
		\includegraphics[width=0.5\columnwidth]{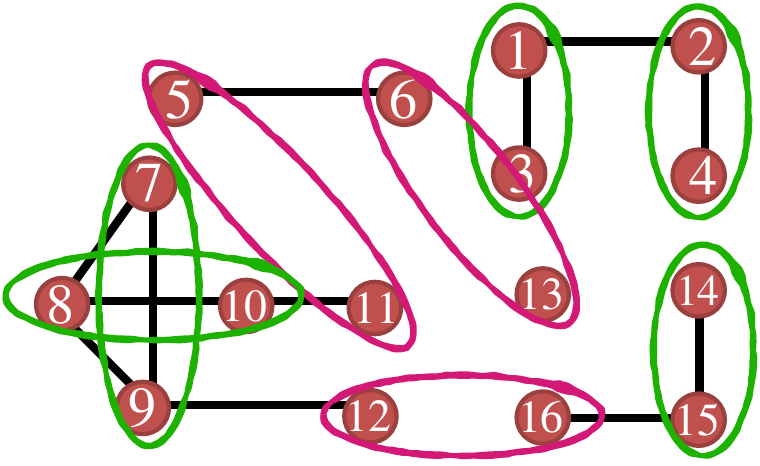}
	}
	\caption{	
	The flipped parity bits are closely spaced.  (a) \methodName{} localy predecode 3 pairs (circled in blue) and sends the rest to the main decoder (b) Astrea-G could not prune the paths among components, resulting in incorrectly matching flipped bits across components. }

	\label{fig:promatch_decodes_AG_fails}
\end{figure}

\methodName{}, on the other hand, excels in these scenarios by prioritizing safer and easier matchings. Its local approach focuses on quickly resolving simpler pairings in the immediate vicinity. The remaining, potentially complicated parts of the syndrome are then managed by the main decoder, within the remaining time until reaching the 1-$\mu s$ threshold. In Figure~\ref{fig:promatch_decodes_AG_fails}, we depict an example of these cases, which components are close to each other but the correct matching does not have any pairs among separate components. In this case, \methodName{} correctly predecodes three pairs, circled in blue in Figure~\ref{fig:promatch_decodes_AG_fails}(a), and sends the rest to the main decoder. On the other hand, as shown in Figure~\ref{fig:promatch_decodes_AG_fails}(b),  Astrea-G's solution contains matching among components due to the closeness of the components which prevent pruning of the MWPM graph.

\textbf{Astrea-G Succeeds, \methodName{} Struggles:} While Astrea-G also employs a greedy algorithm, it shows proficiency in decoding scenarios where some components of the decoding subgraph are distantly placed, facilitating the pruning process and reducing the size of the search space. This is particularly advantageous in cases where certain components are far enough apart to allow for effective pruning, yet others are close enough to provide lower matching solutions overall. Especially, if there exists components with an odd number of parity bits, which necessitate cross-component matchings. Our experiments showed that almost all of samples that \methodName{} fails to decode ($99.9\%$ for \methodName{} compared to $61\%$ for Astrea-G) contains components with odd number of nodes. In such instances, Astrea-G’s ability to balance between pruning distant components and exploring close pairings becomes crucial.

\begin{figure}[h]
	\captionsetup[subfigure]{position=top} 
	\centering	
	\subfloat[]{
		\includegraphics[width=0.4\columnwidth]{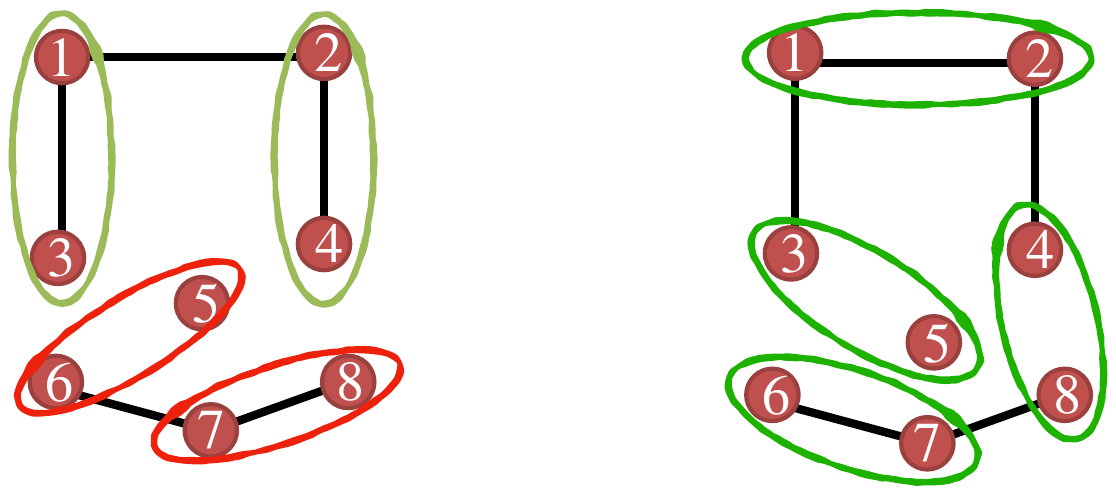}
	}\hspace*{-0.8em}		
	\subfloat[]{
		\includegraphics[width=0.35\columnwidth]{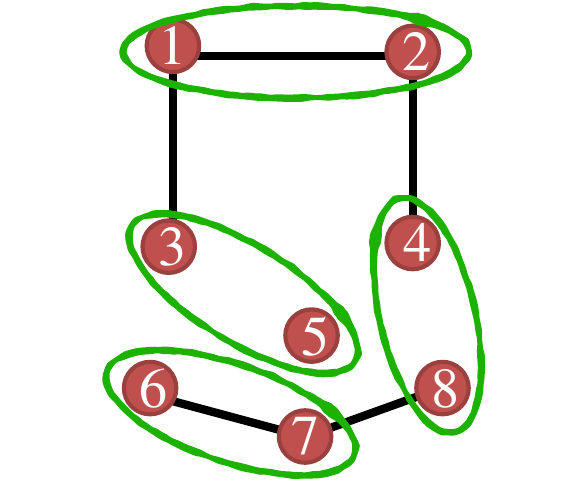}
	}
	\caption{	
	There exist multiple matchings across components (a) \methodName{} predecode two pairs incorrectly, since node 5 is not close enough to any of the remaining nodes (b) Astrea-G near-exhaustive search finds the correct solution.
}	
	\label{fig:promatch_fails_AG_decodes}
\end{figure}

\methodName{}, with its focus on local pairings, may not efficiently decode these error patterns. It excels in scenarios with more localized error structures but struggles when the optimal decoding path involves considering a wider spread of error components (more than 10 flipped bits so that no all of them can be passed to the main decoder). Astrea-G's method of selectively pruning the search space, while still considering broader pairings, allows it to uncover decoding paths that \methodName{} might miss due to its local focus. See Figure~\ref{fig:promatch_fails_AG_decodes} as an example for such scenarios.

\textbf{Both Face Challenges:} Despite their respective strengths, \methodName{} and Astrea-G both encounter difficulties in certain complex scenarios. These challenging situations involve dense error patterns that require exhaustive searches beyond what either algorithm can handle. In these instances, the error patterns are so densely packed that they exceed the search capabilities of both \methodName{} and Astrea-G within the required timeframe. While such scenarios are infrequent, they contribute to a slight increase in the LER ( e.g. $10^{-17}$ for $d=13$). However, this impact on the LER is not significant enough to notably affect the overall decoding performance. The rarity of these complex cases ensures that the effectiveness of parallel running of \methodName{} and Astrea-G in real-time decoding remains high.

\subsection{Comparison of \methodName{} with Prior Predecoders}
In the realm of predecoders for surface codes, various methodologies have been explored. The hierarchical predecoder~\cite{delfosse2020hierarchical} and Clique~\cite{ravi2023better} adopt a greedy strategy, matching each flipped parity bit with its neighbors. It operates on a non-syndrome-modified basis: if it fails to decode all flipped bits, it forwards the entire syndrome unaltered to the main decoder. These predecoders do not adequately address the issue of high Hamming weight syndromes, as they do not modify these complex syndromes before passing them to the main decoder. \methodName{}, in contrast, modifies syndromes in such a way that the main decoder can process them within the remaining time, thus easing the decoding process for high distance codes ($d > 9$).

Neural network-based predecoders, represented by Chamberland et al.~\cite{chamberland2022hierarchical} and NEO-QEC~\cite{ueno2022neoqec}, modify syndromes but are not designed for real-time decoding due to latency issues on FPGAs~\cite{chamberland2022hierarchical} and relying on emerging technologies~\cite{ueno2022neoqec}. \methodName{} offers an advantage over these methods by being optimized for real-time decoding on FPGAs, solving the latency issues inherent in these approaches.

Belief propagation represents a novel approach in predecoding proposed by Caune et al.~\cite{caune2023belief}, targeting high accuracy and coverage. However, its effectiveness relies on the algorithm's ability to converge to a solution. When it does not achieve convergence, the algorithm partially decodes the syndrome, modifies it, and adjusts the weights of the decoding graph before sending the remainder to the main decoder. This process, involving complex message passing among the nodes of the decoding graph, presents significant challenges for real-time implementation. To date, a real-time execution of belief propagation on hardware, particularly on FPGA, has not been demonstrated. \methodName{}, in contrast, offers a simpler and more reliable solution. It employs a straightforward logic that has been effectively implemented on FPGA, ensuring consistent predecoding performance without the complexities of message passing.The qualitative differences among various predecoders are summarized in Table~\ref{tab:predecoders_comparison}, with respect to accuracy, coverage, and real-time (RT) capability.

 \begin{table}[ht]
\centering
\caption{Comparison of \methodName{} with Prior Predecoders}
\label{tab:predecoders_comparison}
    {
    \begin{tabular}{|l |c |c |c |c |}
    \hline
    
        \textbf{Predecoder} & \textbf{Accuracy} & \textbf{Coverage} & \textbf{RT}  \\ \hline\hline
         \methodName{}            & High & Sufficient & Yes         \\ \hline
         Clique~\cite{ravi2023better}$^1$         & High & Low & Yes         \\ \hline
        Hierarchical~\cite{delfosse2020hierarchical}$^1$   & High &  Low  & Yes      \\ \hline
        Smith et al.~\cite{smith2023local}   & Low & High   & Yes      \\ \hline
        Chamberland et al.~\cite{chamberland2022hierarchical} & Low & High  & No \\ \hline
        NEO-QEC~\cite{ueno2022neoqec} & Low & High  & No \\ \hline
        Belief Propagation~\cite{caune2023belief} & High & High  & No \\ \hline
    \end{tabular}}
    \begin{tablenotes}
    \footnotesize
   \item $^1$\textit{Despite having high accuracy, due to very low coverage, the accuracy of final LER of these methods is very low, due to the limitations of RT-MWPM.}
    \end{tablenotes} 
\end{table}

\section{Evaluation Methodology}

\subsection{Surface Code}
We consider rotated surface codes for distance $11$ and $13$. Given that the recent Astrea-G decoder achieves RT-MWPM up to $d = 9$, we seek to achieve RT-MWPM up to $d = 13$.

\subsection{Evaluation Baseline}
As our goal is to demonstrate RT-MWPM up to $d = 13$, we use idealized MWPM as a baseline. The closer a decoder's logical error rate is to MWPM's logical error rate, the better. We also use Astrea-G~\cite{vittal2023astrea} decoder, Clique~\cite{ravi2023better}, and Smith et al. predecoders~\cite{smith2023local}  as other baselines.

\subsection{Noise Model and Simulation Infrastructure\protect\footnotemark}
\footnotetext{The source code of our implementation is available for access at \url{https://github.com/nargesalavi/Promatch}.}
We adopt a uniform circuit-level noise model with a physical error rate ranging from $p = 10^{-4}$ to $5 \times 10^{-4}$. This model includes (1) start-of-round depolarizing errors (with equal probabilities for $I, X, Y, Z$) on data qubits, (2) depolarizing errors following gate operations on all qubit operands, (3) measurement errors, and (4) reset initialization errors, each occurring with probability $p$. The use of a uniform physical error rate is widely acknowledged in quantum-error-correction research, as evidenced by several studies~\cite{vittal2023astrea, maurya2023scaling, gidney2021honeycombcode, das2022afs, ravi2023better, fowler2012surface, chamberland2019fault}. Additionally, employing a circuit-level error model is considered reflective of real-device performance, aligning with the standards set in recent research \cite{google2023suppressing, landahl2011colorcodes, fowler2012surface, gidney2021stim, gidney2022benchmarkinghoneycomb, vittal2023astrea}. 
We use Google's Stim framework~\cite{gidney2021stim} for our evaluations due to its status as an industry-scale simulator, well-established in  quantum error correction studies~\cite{vittal2023astrea, maurya2023scaling,gidney2021honeycombcode}. This ensures the reliability and relevance of our experimental approach. In these state-preservation, or memory, experiments~\cite{das2022lilliput, das2022afs, holmes2020nisqplus, ueno2021qecool, ueno2022qulatis, google2022distance5, sundaresan2022ibmqec, chen2021exponential, ryananderson2021realizationftqc, vittal2023astrea}, we initialize a logical qubit in the $|0\rangle$ state and perform syndrome extraction over $d$ rounds, followed by measurement in the computational basis. The success of each experiment, and the estimation of the decoder's LER, is based on whether the measurement outcome aligns with the decoder's correction. Repeating these experiments across millions of trials allows for an accurate estimation of logical error rates\footnote{In this paper, we use only $Z$ memory experiments, equivalent to $X$ experiments with qubit initialization to $|+\rangle$ and Hadamard basis measurement.}.

To simulate the expected low LER (i.e. in the order of  $10^{-15}$), which would otherwise require trillions of trials, we use an alternate approach~\cite{flair}. For up to $ k = 24$ random error injections, we generate millions of syndromes and calculate the decoding failure probability, $P_f(k)$, and the occurrence probability, $P_o(k)$, of these errors. The logical error rate is then estimated using Equation~(\ref{eq:faultinj}), based on our error model.

\begin{equation}
    \label{eq:faultinj}
    \text{Logical Error Rate} = \sum_{k} P_o(k) \times P_f(k)
\end{equation}

\section{Evaluations}

We evaluate \methodName{} and \methodName{} || AG for $d=11$ and $d=13$, for $p=10^{-4}$ to $5 \times 10^{-4}$.


\subsection{Logical Error Rate of \methodName{}}
Table~\ref{tab:LER} shows the LER of \methodName{}, idealized MWPM, Astrea-G~\cite{vittal2023astrea}, and Smith et al.~\cite{smith2023local} predecoder for $d= 11$ and $d = 13$. Smith et al. predecoder is only applied to high-Hamming weight (HW $>10$) syndromes and uses Astrea as the main decoder (same as \methodName{}). Smith predecoder cannot improve the LER of Astrea beyond $d=11$ due to its low accuracy and not guaranteeing enough coverage. For $d = 11$, \methodName{} gains similar LER to Astrea-G for $p=10^{-4}$. For $d = 13$, \methodName{} outperforms Astrea-G by $5.6\times$. Furthermore, we observe parallel running of \methodName{} and Astrea-G achieves practically identical performance to MWPM, outperforming the setting which Smith is running in parallel to Astrea-G.


\begin{table}[h]
\vspace{-0.05in}
    \caption{Logical Error Rate for $d = 11$ and $d = 13$ at $p = 10^{-4}$}
  \centering
  \begin{adjustbox}{max width=\columnwidth}
\setlength{\tabcolsep}{0.24cm} 
\renewcommand{\arraystretch}{1.4}
    \begin{tabular}{| l | c | c |}
        \hline
         Decoder & \textbf{$d = 11$} & $d = 13$ \\
         \hline \hline
         MWPM (\textit{Ideal}) & $1.8\times 10^{-13}$ & $3.4\times 10^{-15}$ \\ \hline
         \methodName{}$^1$ || AG & $1.8\times 10^{-13}$ ($1\times$) & $3.4\times 10^{-15}$ ($1\times$)\\
         \hline
         \methodName{} + Astrea & $4.5\times 10^{-13}$ ($2.5\times$) & $2.6\times 10^{-14}$ ($7.7\times$) \\ 
         \hline
         Astrea-G (AG) & $4.5\times 10^{-13}$ ($2.5\times$) & $1.4\times10^{-13}$ ($43\times$)\\
         \hline
         Smith$^1$ || AG & $2.5\times 10^{-13}$ ($1.3\times$) & $1.5\times10^{-14}$ ($4.5\times$) \\
          \hline
         Smith + Astrea  & $4.4\times 10^{-11}$ ($240\times$) & $6.9\times 10^{-11}$ ($20412\times$) \\
         \hline
         
    \end{tabular}
    
    \end{adjustbox}
    \label{tab:LER}
     \begin{tablenotes}
     \footnotesize
    \item $^1$\textit{The main decoder in this setting is Astrea. In other words, structure of this design is (predecoder + Astrea) || AG}
    \end{tablenotes} 
\end{table}

We also experimented with the  Clique~\cite{ravi2023better} predecoder in two settings:~1)~Astrea as the main decoder (Clique + Astrea), 2) Astrea-G as the main decoder (Clique+AG). In both cases, the predecoder is only applied on High-Hamming weight syndromes, as syndromes with low Hamming weights can perfectly be decoded by the main decoder (same as \methodName{}). 
Table~\ref{tab:LER_clique} shows the LER for Clique. Clique+Astrea has a high LER as Clique is a non-modified predecoder that does not handle complex patterns. As distance increases, the probability of encountering complex patterns, particularly at high-Hamming weight syndromes (HW $>10$) increases. In these instances, Clique forwards these syndromes to Astrea. Astrea cannot decode any of them because it is optimized for decoding only low-HW (HW $\leq 10$) syndromes (this results in the LER of Clique + Astrea becomes too high, in the order of $p$, for distance $13$). The LER of Clique+AG is equal to the LER of AG. This shows Clique cannot improve the performance of main decoders due to its too low coverage. Therefore, we do not analyze Clique any further in our study.

\begin{table}[h]
\vspace{-0.05in}
    \caption{Clique's Logical Error Rate for $d = 11$ and $d = 13$, and $p = 10^{-4}$.}
  \centering
  \begin{adjustbox}{max width=\columnwidth}
\setlength{\tabcolsep}{0.24cm} 
\renewcommand{\arraystretch}{1.3}
    \begin{tabular}{| l | c | c |}
        \hline
         Decoder & \textbf{$d = 11$} & $d = 13$ \\
         \hline \hline
         Clique + Astrea  & $2.2\times 10^{-5}$ ($10^{8}\times$) & $> 10^{-4} (> 10^9 \times)$\\
         
         \hline
         Clique + AG  & $4.5\times 10^{-13}$ ($2.5\times$) & $1.4\times10^{-13}$ ($43\times$) \\
         
         \hline
         Astrea-G (AG) & $4.5\times 10^{-13}$ ($2.5\times$) & $1.4\times10^{-13}$ ($43\times$)\\
         \hline
    \end{tabular}
    
    \end{adjustbox}
    \label{tab:LER_clique}
\end{table}
\subsection{Sensitivity to Physical Error Rate}
\begin{figure}[h!]
	\centering	
	\includegraphics[width=\columnwidth]{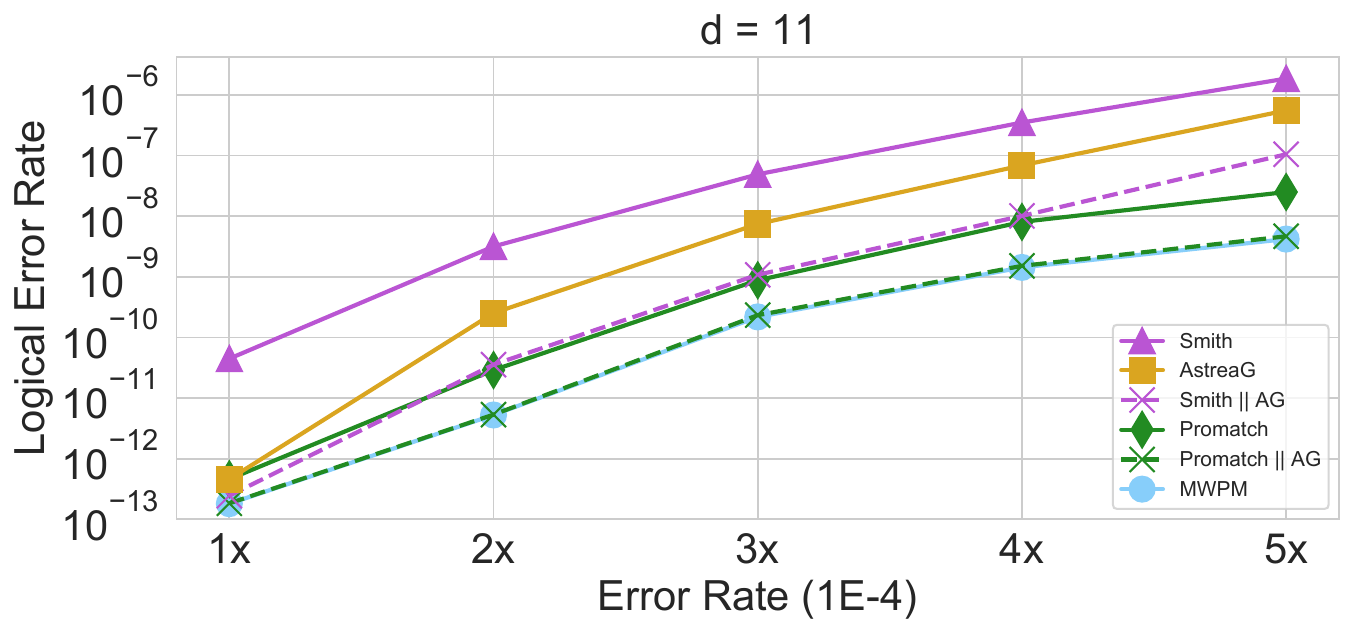}
    \caption{LER of non-real-time MWPM, Promatch, AstreaG(AG), Smith, Smith~||~AG, Promatch || AG for $10^{-4} \leq p \leq 5 \times 10^{-4}$ for $d=13$. \methodName{}~||~AG remains in $1.1\times$ of MWPM's LER.}
	\label{fig:sensitivty_d11}
\end{figure}

\begin{figure}[h!]
	\centering	
	\includegraphics[width=\columnwidth]{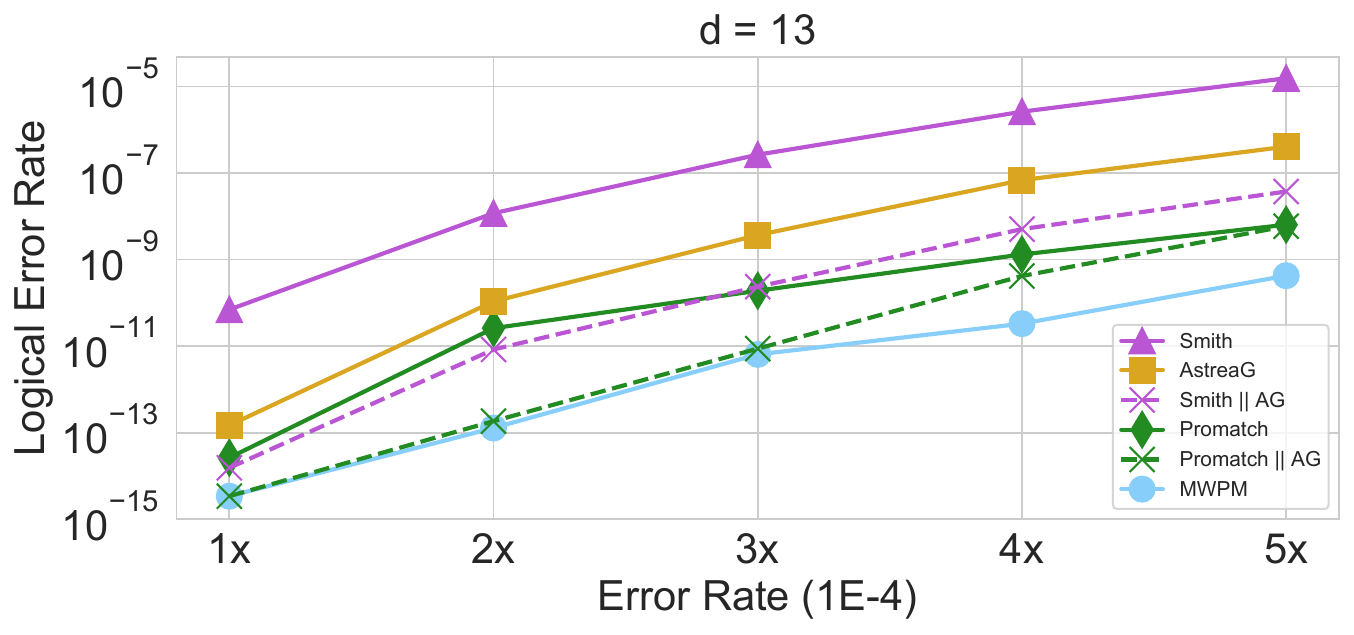}
    \caption{LER of non-real-time MWPM, Promatch, AstreaG(AG), Smith, Smith~||~AG, Promatch || AG for $10^{-4} \leq p \leq 5 \times 10^{-4}$ for $d=13$. \methodName{}~||~AG remains in $13.9\times$ of MWPM's LER.}
	\label{fig:sensitivty_d13}
\end{figure}

Figures~\ref{fig:sensitivty_d11} and~\ref{fig:sensitivty_d13} present the LER for idealized MWPM, \methodName{}, Astrea-G (AG), Smith, Smith~||~AG, and \methodName{}~||~AG over physical error rates from $10^{-4}$ to $5 \times 10^{-4}$. \methodName{} maintains an LER within $5.9 \times$ to $202 \times$ of MWPM's LER for $d = 11$ and $13$, respectively. This performance significantly outperforms Astrea-G, whose LER is up to $22 \times$ and $2064 \times$ higher than MWPM's for distance 11 and 13, respectively. \methodName{}~||~AG remains in $1.1\times$ and $ 13.9 \times$ of MWPM's LER for distance 11 and 13, respectively, significantly outperforms Smith~||~AG which remains in $25 \times$ and $152 \times$ of MWPM's LER for distance 11 and 13, respectively.

\subsection{Hamming Weight Reduction of Syndromes}

\begin{figure}[h]
	\centering	
	\includegraphics[width=\columnwidth]{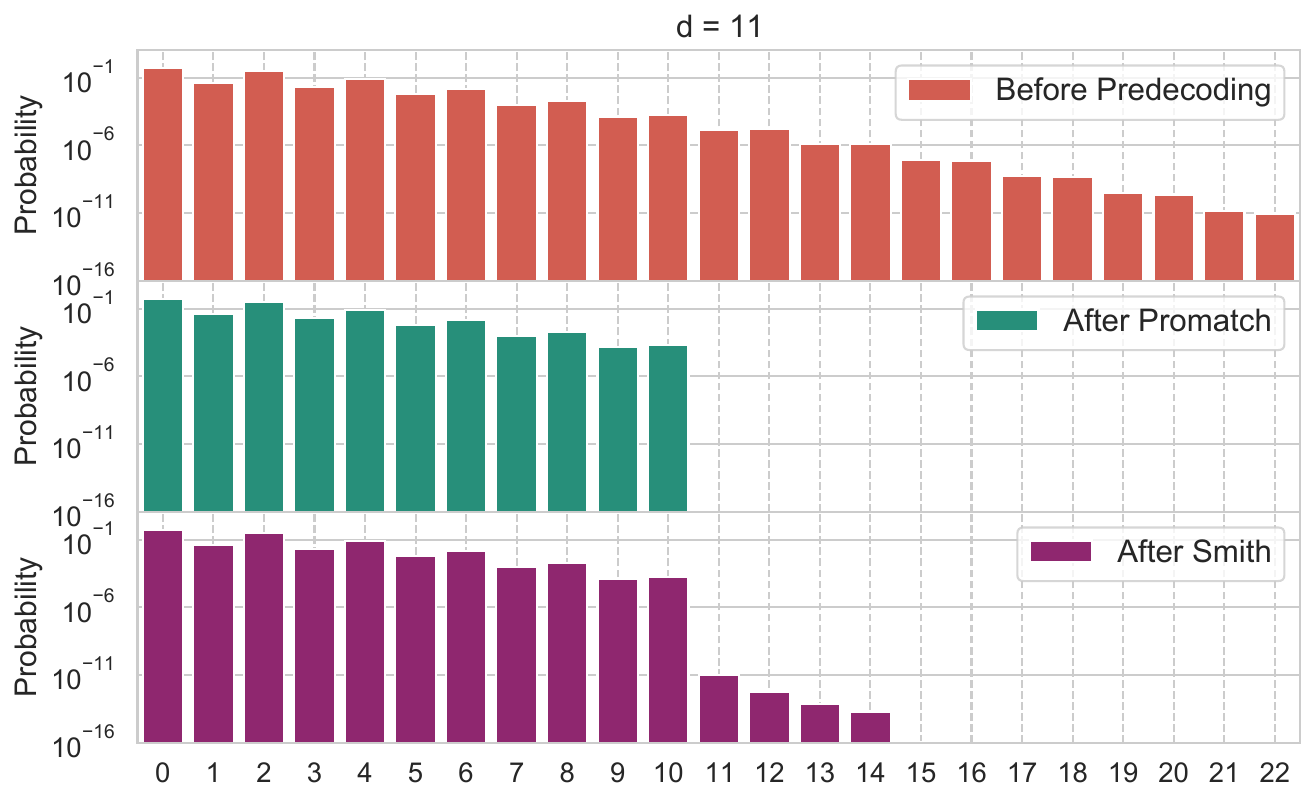}
    \caption{\methodName{} consistently lowers syndrome Hamming weight to 10 or less, allowing Astrea to accurately decode distance 11 surface codes, unlike the Smith et al. predecoder, for $p = 10^{-4}$}
	\label{fig:hw_distr_11}
\end{figure}

\begin{figure}[h]
	\centering	
	\includegraphics[width=\columnwidth]{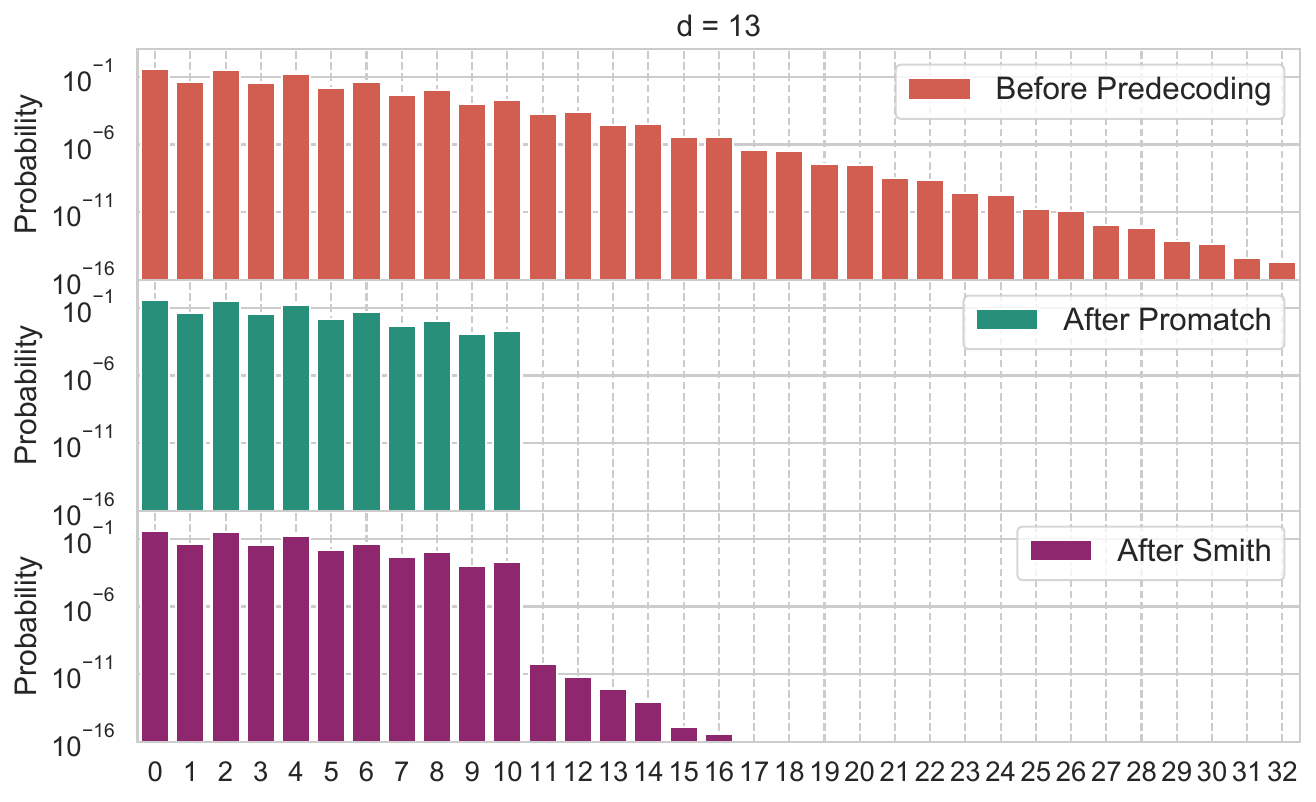}
    \caption{Unlike Smith et al. predecoder, \methodName{} reduces the syndrome Hamming weight up to 10 such that the main decoder (Astrea) can decode distance 13 surface codes accurately for $p = 10^{-4}$.}
	\label{fig:hw_distr_13}
\end{figure}

Figures~\ref{fig:hw_distr_11} and~\ref{fig:hw_distr_13} show the HW distribution before and after applying syndrome-modified predecoding methods, namely \methodName{}, and Smith et al.~\cite{smith2023local} for $d=11$ and $d=13$. In our experiments, the predecoding only applies to high-HW syndromes ($HW > 10$), as Astrea~\cite{vittal2023astrea} can decode low-HW syndromes in real-time. Unlike Smith et al~\cite{smith2023local}, \methodName{} always predecodes high-HW down to HWs of 6, 8, or 10. \methodName{} consistently achieves sufficient coverage, ensuring Astrea can decode all post-predecoding syndromes.

\subsection{Latency of \methodName{} and Its Impact on LER}

The latency of \methodName{} depends on the number of edges in the decoding subgraph since the pipeline iterates multiple times over these edges. During each cycle, \methodName{}'s pipeline analyzes one edge of the subgraph and decides whether or not to match the syndrome bits connected by this edge. If two syndrome bits are matched, the edges linked to the recently matched bits are removed from the decoding graph. Consequently, in the next predecoding round of that syndrome, the \methodName{} pipeline operates over fewer edges.

Based on this pipeline structure, we estimated the number of consumed cycles for each syndrome by summing the edge numbers in the decoding subgraphs across all predecoding rounds prior to sending the syndrome to the main decoder. When utilizing Step 3 of \methodName{}, we add the maximum value of the following: the number of paths from singletons to flipped bits or the number of edges.

If the duration exceeds $1\mu$s, it is categorized as a logical error, prompting an abort of \methodName{}. We have allocated $10$ cycles for the final comparison of the solution with Astrea-G~\cite{vittal2023astrea} in the \methodName{}~||~AG design. Thus, in our simulations, \methodName{}, inclusive of the main decoder, is allotted a time budget of $960ns$ (operating at 250MHz).

Tables \ref{tab:decoder_latency} and  \ref{tab:pred_latency} depict the maximal and average latencies for predecoding and the combined predecoding + main decoder, respectively. It's crucial to note that all provided numbers pertain to decoding high Hamming weight syndromes where HW $\geq 10$. Is important to note that in Tables~\ref{tab:pred_latency}, the maximum latency shows that there are cases that \methodName{} reaches its maximum budget, while there is a very low probability $1.5\times 10^{-17}$ for $d=13$ which \methodName{} exceeds $1\mu  s$, negligibly impacts the LER. Moreover, as \methodName{} is a lightweight predecoder, one can run multiple number of its pipeline in parallel for reducing the latency further.
\begin{table}[h!]
    \caption{Latency of Predecoding High-HW Syndromes ($ns$)}
    \centering
    \setlength{\tabcolsep}{0.9cm} 
    \renewcommand{\arraystretch}{1.5} 
    \small 
    \begin{tabular}{| l  |c  |c  |}
        \hline
        Distance & 11 & 13 \\
        \hline \hline
        Max & $824$ &  $928$ \\ \hline
        Average & $68.2$ & $70.0$ \\
        \hline
    \end{tabular}
    \label{tab:decoder_latency}
\end{table}

\begin{table}[h!]
    \caption{Latency of Decoding High-HW Syndromes using \methodName{} ($ns$)}
    \centering
    \setlength{\tabcolsep}{0.8cm} 
    \renewcommand{\arraystretch}{1.5} 
    \small 
    \begin{tabular}{| l | c | c | }
        \hline
        Distance & 11 & 13 \\
        \hline \hline
        Max & $904$ &  $960$ \\ \hline
        Average & $524.2$ & $526.0$ \\
        \hline
    \end{tabular}
    \label{tab:pred_latency}
\end{table}

\subsection{Usage of Each Steps of \methodName{}}
Different steps of \methodName{} is utilized with different rates because of different probabilities of Hamming weights in the syndrome and different rate of simple and complex patterns. 
Table~\ref{tab:steps} illustrates the proportion of samples, for d=11 and d=13, that are processed up to each of four distinct steps in the algorithm. The value associated with each step signifies the frequencies of samples that needed to be processed up to that point. It shows an overview of the efficiency of each step of \methodName{}, reflecting how often each step is necessary to reach the final LER. For $d = 11$ and $13$, $99.56\%$ and $99.83\%$ of samples require only Step 1 of the algorithm, respectively. Nonetheless, other steps of the \methodName{} play an important role in the predecoding process due to the very low LER values. The last step, which is utilized with the least frequency, is still employed with higher frequency than LER for all the distances, which shows the importance of all the steps of the algorithm in reaching the desired LER. 

\begin{table}[h]
    \caption{Frequency of each step during the decoding process}
  \centering
\setlength{\tabcolsep}{0.6cm} 
\renewcommand{\arraystretch}{1.6}
    \begin{tabular}{|  l | c | c  |}
        \hline
         Steps & d = 11 & d = 13 \\
         \hline \hline
         Step 1 & $0.9956$ & $0.9983$ \\ \hline
         Step 2 & $0.00439$ & $0.00167$ \\ \hline
         Step 3 & $6.1\times 10^{-11}$ & $7.3\times 10^{-11}$ \\ \hline 
         Step 4 & $2.4\times 10^{-11}$ & $1.8\times10^{-11}$ \\
         \hline
    \end{tabular}
    \label{tab:steps}
\end{table}
\vspace*{0.35cm}
\subsection{FPGA Utilization and Storage Overhead of \methodName{}}

 {We synthesize \methodName{} on a Kintex UltraScale+ FPGA. The utilization details are shown in Table~\ref{tab:fpga_utilization}. \methodName{}'s efficient use of resources underscores \methodName{}'s practicality for near-term real-time decoding.}

\begin{table}[ht]
    \centering
    \setlength{\tabcolsep}{0.25cm}
    \caption{FPGA Utilization of \methodName{}}
    \label{tab:fpga_utilization}
    
    {

    \begin{tabular}{|l|c|c|c|}
        \hline
        Resource & LUT & FF & Frequency \\
        \hline
        \hline
        Edge-Processing Pipeline & 3\% & 1\% & 250 MHz \\
        \hline
    \end{tabular}}
\end{table}

Table~\ref{tab:storage_requirements} shows the memory needed for storing the Edge table and the Path table of \methodName{}. For storing the Path Table, we optimize the required memory by categorizing the paths into four groups as Promatch is not sensitive to the exact weight of the paths.

\begin{table}[ht]
    \centering
    \setlength{\tabcolsep}{0.7cm}
    \caption{Storage Requirements of \methodName{}}
    \label{tab:storage_requirements}
    {
    \begin{tabular}{|l|c|c|}
        \hline
        Storage Type & d = 11 & d = 13 \\
        \hline
        \hline
        Edge Table & 3.6 KB & 6 KB \\
        \hline
        Path Table & 129 KB & 345 KB \\
        \hline
    \end{tabular}}
\end{table}





\section{Related Work}

We discuss prior work on real-time decoding and compare and contrast them with \methodName{}.

\subsection{Decoders for Surface Code}

Error decoding has been an active area of research, with several decoding algorithms proposed in the literature. The different classes of decoding algorithms:

\noindent \textbf{Lookup Table (LUT) Decoder}: 
This method uses a lookup table (LUT) to store corrections for every syndrome. Typical LUT implementations has scalability challenges due to exponential storage overheads.

\vspace{0.05in}

\noindent \textbf{Minimum Weight Perfect Matching (MWPM)}: This decoder uses a graph pairing algorithm and is considered to be one of the most effective in terms of accuracy. There has been significant recent research to improve the latency and scalability of MWPM especially for smaller distance codes.

\vspace{0.05 in}
\noindent \textbf{Machine Learning (ML) Decoders}: These designs train neural networks with a set of syndrome vectors and the resulting error location and types~\cite{breuckmann2018scalable,torlai2017neural,baireuther2018machine,krastanov2017deep,varsamopoulos2017decoding,chamberland2018deep,maskara2019advantages,sweke2018reinforcement,baireuther2019neural,ni2018neural,andreasson2019quantum,davaasuren2018general,liu2019neural,varsamopoulos2018designing,wagner2019symmetries,chinni2019neural,colomer2019reinforcement}. They key challenges with these designs is the lack of training data for large distance codes and the substantial resource requirements (memory, compute, time), making them unappealing for large distance codes.

\subsection{Real-Time Decoding}

Real-time decoding is necessary to achieve quantum fault-tolerance, and thus much prior work has been dedicated to achieving accurate decoding in real-time. Recent implementations of MWPM decoding, such as \textbf{Sparse-Blossom} and \textbf{Fusion-Blossom}~\cite{higgott2023pymatching2, wu2023fusion}, have achieved \textit{mean} latencies of $1\mu$s per syndrome extraction round. While such results are impressive, quantum fault-tolerance will likely require a worst case latency of $1\mu$s to avoid unnecessarily stalling the quantum computer, and thus such implementations are insufficient for real-time decoding. Consequently, much recent work has proposed hardware implementations. \textbf{LILLIPUT} is capable of decoding up to $d = 5$ with the same accuracy as MWPM~\cite{das2022lilliput}. However, the lookup tables of LILLIPUT are difficult to scale beyond $d = 5$. Decoders such as \textbf{NISQ+} and \textbf{QECOOL}/\textbf{QULATIS} leverage superconducting logic to achieve low latencies~\cite{holmes2020nisqplus, ueno2021qecool, ueno2022qulatis}. However, these decoders cannot handle arbitrary measurement errors, thus causing significant inaccuracy. The \textbf{AFS} decoder implements the union-find algorithm~\cite{das2022afs}. AFS is comparable to MWPM at extremely low error rates, but for the near-term error rate of $10^{-4}$ evaluated in this paper, AFS is inaccurate compared to MWPM decoding as the underlying union-find algorithm is inaccurate in this regime~\cite{delfosse2017unionfind}. \textbf{Astrea} is a recent implementation of MWPM decoding restricted to distances $7$ and $9$, but has poor performance for $d = 11$ and beyond~\cite{vittal2023astrea}. \methodName{} achieves real-time decoding up to $d = 13$ with comparable accuracy to MWPM and is the first decoder to do so.

\subsection{Predecoding for Reduced Bandwidth}

Predecoding has emerged as a strategy for (1)~reducing bandwidth requirements and (2)~improving the latency of decoders. Delfosse~\cite{delfosse2020hierarchical}  demonstrated that predecoding can offer significant bandwdith reduction without sacrificing accuracy. Recently, the Clique decoder provided an implementation of Delfosse's design in superconducting logic~\cite{ravi2023better}. However, while both Delfosse's proposal and the Clique decoder reduce bandwidth requirements, they do not improve worst-case decoder latency. Smith et al. has proposed a predecoder which greedily filters syndromes sent to the main decoder, allowing for better coverage compared to Delfosse's original proposal~\cite{smith2023local}. Concurrently, Chamberland et al. proposed a predecoder to filter syndromes, but uses neural networks instead of a greedy strategy~\cite{chamberland2022hierarchical}; NEO-QEC is an implementation of Chamberland et al.'s proposal in superconducting logic~\cite{ueno2022neoqec}. However, the limitation of these works is that they sacrifice accuracy to improve coverage. In contrast, \methodName{} adaptively finds  the highest accuracy at a good-enough coverage to ensure real-time decoding.

\section{Conclusion}

Decoders used in quantum error correction must accurately identify errors in real-time (typically within a $1 \mu s$ on superconducting systems) to prevent the backlog of errors. Although the Minimum Weight  Perfect Matching (MWPM) decoder is widely recognized for its effectiveness in decoding surface codes, achieving MWPM accuracy  beyond distance 9 remains an open problem because the complexity of decoding grows with the distance of the code. This paper introduces \methodName{} to expand the reach of real-time MWPM up to distance 13. \methodName{} uses predecoding to transform high Hamming-weight syndromes into low Hamming-weight that are amenable to real-time MWPM decoders. 

Designing an accurate yet high-coverage predecoder is challenging.  Aggressive predecoding leads to inaccurate matches, whereas conservative predecoding does not reduce the complexity of the decoding graph substantially and remains bottlenecked by the complexity of the main decoder. \methodName{} attains a sweet-spot between these two extremes by leveraging the following  insights. First, most syndrome bit flips are matched to other flipped syndrome bits in their local neighborhood. Second, predecoding must be performed only until the syndrome Hamming weight reaches a point beyond which it can be fully handled by the MWPM decoder. \methodName{} enables locality-aware adaptive predecoding that adjusts the number of prematches depending on the syndrome Hamming weight to ensure high accuracy while simultaneously meeting the real-time latency constraints. \methodName{} achieves logical error rate of $4.5 \times 10^{-13}$ and $2.6 \times 10^{-14}$ for distance 11 and 13, respectively. \methodName{} also achieves MWPM accuracy for up to distance 13 when it runs in parallel with Astrea-G (logical error rate of $3.4 \times 10^{-15}$). 

\section*{Acknowledgement}
We thank the reviewers of ASPLOS-2024 for their suggestions and feedback. N.A. thanks Sadegh Shirani for helpful discussions.  This work was
funded in part by EPiQC, an NSF Expedition in Computing,
under grant CCF-173044.

\bibliographystyle{plain}
\bibliography{references}


\end{document}